\documentclass[journal]{IEEEtran}
\usepackage{times}
\usepackage{amsmath,amsfonts,amssymb}
\usepackage{algorithmic}
\usepackage{algorithm}
\usepackage{array}
\usepackage{graphicx}
\usepackage{textcomp}
\usepackage{stfloats}
\usepackage{url}
\usepackage{verbatim}
\usepackage{xcolor}
\usepackage{tabularx}
\usepackage{booktabs}
\usepackage{threeparttable}
\usepackage{multirow}
\usepackage{bm}
\usepackage{cite}
\usepackage[table]{xcolor}
\usepackage[colorlinks=true,linkcolor=blue,citecolor=blue,urlcolor=blue]{hyperref}
\usepackage[caption=false,font=normalsize]{subfig}

\makeatletter
\renewcommand{\footnoterule}{%
  \kern -6pt
  \hrule \@width \columnwidth \@height 0.4pt
  \kern 5pt
}
\makeatother
\begin{document}

\title{AaSP: Aliasing-aware Self-Supervised Pre-Training\\for Audio Spectrogram Transformers}

\author{Kohei~Yamamoto$^{1,2}$, Kosuke~Okusa$^{2}$
\thanks{$^{1}$Research \& Development Center, Technology Division, Oki Electric Industry Co., Ltd., Saitama, Japan. $^{2}$Department of Data Science for Business Innovation, Chuo University, Tokyo, Japan. Kohei~Yamamoto (e-mail: yamamoto833@oki.com) is a Ph.D. student at Chuo University and also with Oki Electric Industry Co., Ltd. Kosuke~Okusa (e-mail: okusa@kc.chuo-u.ac.jp) is an Associate Professor at Chuo University, Tokyo, Japan.}%
}


\maketitle

\begin{abstract}
Transformer-based audio self-supervised learning (SSL) models commonly use spectrograms, vision-style Transformers, and masked modeling objectives. However, convolutional patchification with temporal downsampling lowers the effective Nyquist frequency and introduces aliasing, while naïve low-pass filtering may remove task-relevant high-frequency cues. We present AaSP, an aliasing-aware self-supervised pre-training framework for audio spectrogram transformers. AaSP combines an aliasing-aware patch representation, teacher-student masked modeling, a cross-attention predictor, and multi-mask contrastive regularization to learn representations that integrate features from alias-prone modulation bands while remaining stable across masked views. Its patch-embedding module, Aliasing-aware Patch Embedding (AaPE), augments standard patch tokens with features from alias-prone modulation bands using a band-limited complex sinusoidal kernel with a two-sided exponential window. The kernel’s frequency and decay parameters are estimated from the input, enabling adaptive subband analysis whose outputs are fused with standard patch tokens. We pre-train on AudioSet and evaluate the learned representations by fine-tuning and linear evaluation on acoustic/environmental, speech, and music recognition benchmarks. Under fine-tuning, the full AaSP framework achieves state-of-the-art results on AS-20K, ESC-50, and NSynth among compared self-supervised baselines, while remaining competitive elsewhere. Linear evaluation shows a similar trend, including gains on US8K and NSynth. Overall, AaSP learns representations that are more stable under aliasing-sensitive temporal perturbations and competitive for downstream transfer.
\end{abstract}


\section{Introduction}
\IEEEPARstart{R}{E}CENT advances in natural language processing (NLP) and computer vision demonstrate the effectiveness of self-supervised learning (SSL), thereby training neural networks from unlabeled data via auxiliary objectives.
SSL has achieved strong results in NLP~\cite{bert-19} and vision~\cite{dinov2-23}, and currently is advancing audio research~\cite{audiomae-22, mae-ast-22, ssast-22, atst-24, m2d-24}, improving downstream tasks, such as acoustic recognition, event detection, and captioning.
In the domain of audio, transformer models operating on spectrograms have demonstrated efficacy in learning high-level semantic representations~\cite{ast-21}.
Representing audio as spectrogram-based two-dimensional time-frequency features makes it natural to adapt vision-style architectures such as the Vision Transformer (ViT)~\cite{vit-21} and Masked Image Modeling (MIM)~\cite{mae-22} methods to audio, broadening the scope of acoustic representation learning.
Several early masked/patch-based self-supervised approaches on spectrograms include SSAST~\cite{ssast-22}, which extended the Audio Spectrogram Transformer (AST~\cite{ast-21}) to self-supervised pre-training, and MAE-AST~\cite{mae-ast-22}, which explored MAE-style objectives for audio.
Audio-MAE~\cite{audiomae-22} and subsequent approaches further developed this direction.

Despite the efficacy of Transformer features, a key limitation lies in how spectrograms are patchified for input embeddings.
Many spectrogram-based Transformer pipelines patchify the spectrogram using a convolutional patch embedding with a temporal patch size $P_\text{time}$ and temporal stride $S_\text{time}$ (e.g., AST~\cite{ast-21}).
Depending on the configuration, patches may be non-overlapping ($S_\text{time}=P_\text{time}$) or overlapping ($S_\text{time}<P_\text{time}$), as in AST-style patch embeddings.
In both cases, patchification produces one patch embedding every $S_\text{time}$ spectrogram frames, i.e., the resulting patch sequence is defined on a coarser temporal grid than the input spectrogram when $S_\text{time}>1$.
With a commonly used non-overlapping setting ($S_\text{time}=P_\text{time}=16$) and a 10-ms spectrogram hop, this corresponds to an effective temporal resolution of 160 ms per patch.
This reduction in temporal sampling rate lowers the temporal Nyquist frequency of each subband time series, i.e., the highest temporal modulation frequency representable at the patch rate.
When patch embedding performs temporal downsampling (e.g., via a strided convolution with $S_\text{time}>1$) without explicit anti-aliasing constraints, modulation components above this limit can alias into lower apparent modulation frequencies in the patch-level representation.
In common ViT-style implementations, the patch embedding is realized as a strided convolution that acts as a linear patch projection and is not explicitly constrained to perform anti-aliasing filtering.
We therefore treat aliasing as a potential issue arising from the patch/stride configuration and motivate patch embeddings that incorporate features derived from alias-prone modulation bands without relying on a blanket low-pass filtering strategy that may discard task-relevant high-frequency content.
Aliasing can degrade learned representations and can make outputs sensitive to small input phase shifts~\cite{anti-aliasing-20,improbingsec-21,blurpool-19}.
This can cause instability during training and reduce sample efficiency.
A straightforward solution to this problem is to apply a low-pass anti-aliasing filter prior to downsampling.
However, this approach can discard high-frequency components that are informative for certain tasks.
This creates a fundamental dilemma: we would like to preserve information from modulation components that become vulnerable under patchification, without resorting to blanket low-pass filtering that may discard informative high-frequency content.

To address this issue at the system level, we propose AaSP, an aliasing-aware self-supervised pre-training framework for audio spectrogram transformers.
Its key patch-embedding component is Aliasing-aware Patch Embedding (AaPE), implemented via a Structured Bilateral Laplace Unit (SBLU) inspired by structured state-space models (SSMs)~\cite{s4-22,s4d-22}.
SBLU performs adaptive frequency analysis over spectrogram subbands, focusing on the ranges most susceptible to aliasing under patching.
AaPE extracts features derived from alias-prone modulation bands and fuses them with standard patch tokens to enrich the patch-level representation with information that may otherwise be distorted or underused after strided patchification.
Within AaSP, AaPE derives additional features from alias-prone modulation bands using a band-limited complex sinusoidal kernel with a two-sided exponential window.
This design enables adaptive subband-wise analysis with input-dependent decay and frequency parameters estimated from standard patch tokens.
We describe the detailed formulation of this component in Sec. \ref{sec:sblu} and Sec. \ref{sec:aape}, but emphasize here that its empirical value should be understood as part of the full AaSP design rather than as an isolated replacement module.

\begin{figure*}[t]
  \begin{center}
    \includegraphics[width=0.95\textwidth]{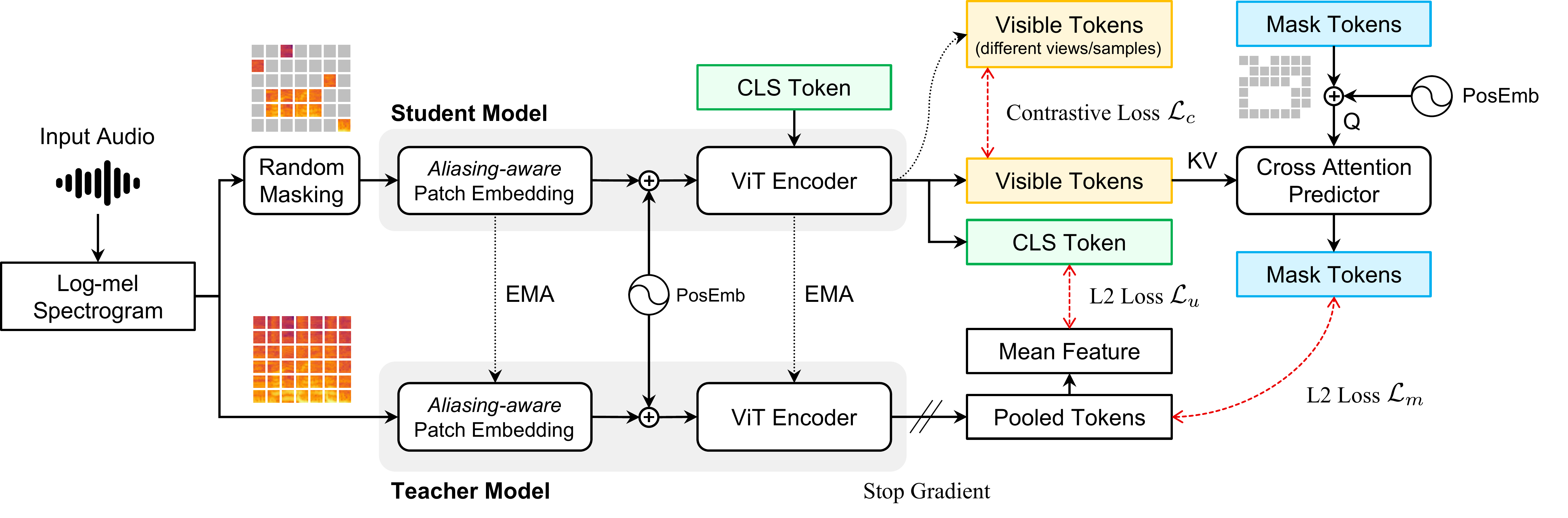}
    \caption{
    \textbf{Overview of the AaSP Framework.} AaSP incorporates \textit{Aliasing-aware} Patch Embedding (AaPE) as its patch-embedding component, where dynamic subband frequency analysis is used to derive aliasing-aware features from aliasing-prone bands and to fuse them with the standard patch tokens. The class (CLS) token predicts a global summary of the teacher outputs, while a cross-attention predictor performs masked prediction of the teacher’s layer-wise pooled tokens at masked positions. A multi-mask contrastive consistency regularization pulls together masked views of the same input and pushes apart views from different inputs. PosEmb denotes absolute positional embeddings\cite{vit-21}.
    }
    \label{fig:overview}
    \vspace{-18pt}
  \end{center}
\end{figure*}

As illustrated in Fig.\ref{fig:overview}, we propose AaSP, an aliasing-aware self-supervised audio pre-training framework.
AaSP builds on an EAT-style masked-modeling backbone~\cite{eat-24} and introduces two main elements: (1) AaPE, which augments patch representations with features derived from alias-prone modulation bands, and (2) a learning design that combines a cross-attention predictor with contrastive regularization across multiple mask patterns.
We evaluate the full framework as a system, as our ablations show that the gains arise from the interaction between the patch-embedding design and the surrounding learning mechanism rather than from patch-embedding replacement alone.
Masked prediction requires the inference of content that may depend on alias-prone modulation bands, which we hypothesize encourages the model to make use of finer-grained cues than conventional approaches.
Furthermore, a multi-mask strategy combined with contrastive regularization encourages consistency across diverse mask patterns, which may facilitate training stability.
We pre-train on AudioSet~\cite{audioset-17} and evaluate via fine-tuning and linear evaluation on downstream benchmarks spanning three broad categories.
(i) General acoustic/environmental sound: AudioSet (AS-2M/AS-20K)~\cite{audioset-17}, ESC-50~\cite{esc50-15}, and UrbanSound8K~\cite{us8k-14}.
(ii) Speech-related: Speech Commands V2~\cite{spcv2-18}, CREMA-D (speech emotion recognition)~\cite{cremad-14}, VoxCeleb1 (speaker identification)~\cite{voxceleb1-17}, and VoxForge (language identification)~\cite{voxforge-18}.
(iii) Music-related: NSynth (instrument family classification)~\cite{nsynth-17}, GTZAN (music genre classification)~\cite{gtzan-02}, and Surge (pitch audio classification)~\cite{surge-21}.
Our results are strongest on general acoustic and music benchmarks, while showing mixed but competitive performance on speech-related tasks.

\textbf{Main contributions.}\quad The contributions of this study are as follows:
\begin{itemize}
        \item Introduce AaSP, an aliasing-aware self-supervised pre-training framework for audio spectrogram transformers, in which AaPE serves as a key architectural component for incorporating features derived from alias-prone modulation bands.
        \item Propose the Structured Bilateral Laplace Unit (SBLU), derived from a structured SSM formulation with a two-sided exponential window, as the mechanism used in AaPE for adaptive subband-wise feature extraction.
        \item Show through ablations that AaPE is not a universally effective standalone replacement, and that its benefit emerges when combined with the surrounding learning design, including multi-mask learning and contrastive regularization.
        \item Provide broad transfer evaluations and additional stability analysis showing that the full AaSP framework yields competitive downstream performance and improved robustness to small temporal shifts.
\end{itemize}
\section{Related Work}
\subsection{Audio Representation Learning}
Audio representation learning can be broadly categorized into two families: time-domain (waveform) methods and time-frequency (spectrogram) methods.
In both families, masked audio modeling (MAM) has become mainstream.
The input is split into patches, a random subset is masked, and the model reconstructs these regions from visible patches to learn semantic features without
labels.

\textbf{Spectrogram-based SSL.}\quad This line of work applies MAM to spectrograms derived from audio.
For instance, SSAST~\cite{ssast-22}, MAE-AST~\cite{mae-ast-22}, Audio-MAE~\cite{audiomae-22}, and MaskSpec~\cite{maskspec-23} introduce self-supervised learning based on spectrogram reconstruction using AST~\cite{ast-21} as a transformer backbone.
In contrast, BEATs\cite{beats-23} predicts discrete labels at the patch level and relies on teacher-model distillation rather than spectrogram reconstruction.
\par A complementary direction employs self-distillation with an exponential moving average (EMA) teacher formed from the student's parameters and minimizes the prediction error between student outputs and teacher latent features.
This approach can be integrated within the MAM framework.
data2vec~\cite{data2vec-22} and its efficient variant data2vec2.0~\cite{data2vec2-23} define a modality-agnostic teacher-student self-distillation paradigm in which a masked student regresses contextualized EMA-teacher representations of the full or unmasked input instead of modality-specific targets.
EH-MAM~\cite{eh-mam-24} extends this line in speech SSL by introducing teacher-guided easy-to-hard masking based on predicted frame-level reconstruction difficulty.
ATST~\cite{atst-24} feeds augmented visible patches to the student while the EMA teacher receives non-augmented inputs; M2D~\cite{m2d-24} additionally masks the teacher and enforces mutually exclusive masking patterns between student and teacher; ASiT\cite{asit-24} jointly learns spectrogram reconstruction and contrastive teacher-student distillation; EAT~\cite{eat-24} introduces the utterance-frame objective (UFO), which combines a frame-level loss with an utterance-level (i.e., clip-level) loss; the utterance-level term aligns the encoder’s class token with an aggregation of teacher features; ASDA\cite{asda-25} incorporates differential attention within the transformer encoder under the EAT framework.
MATPAC~\cite{matpac-25} is another related EMA teacher-student masked-audio pre-training method, which aligns student predictions to teacher targets using a cross-entropy objective to encourage higher-level representations.
\par By contrast, MAST~\cite{mast-23} and SLICER~\cite{slicer-23} also use momentum-teacher training in audio SSL, but their primary formulations focus on contrastive alignment, multiscale modeling, or clustering-oriented objectives rather than masked teacher-target prediction in the same sense as the above MAM-style methods.
Finally, SSLAM~\cite{sslam-25} adds a loss that brings latent features of spectrogram mixtures closer to the mean of the corresponding un-mixed features.
These models achieve strong results across diverse audio tasks; however, many ViT-style spectrogram methods rely on patchification with relatively coarse temporal strides (e.g., 160 ms per patch token in a common configuration with a 10-ms hop and $P_\text{time}=S_\text{time}=16$), which can exacerbate aliasing-induced information loss.
Notable exceptions exist: for example, ATST~\cite{atst-24} can operate on a finer temporal resolution (on the order of tens of milliseconds per patch token).

\textbf{Waveform-based SSL.}\quad wav2vec 2.0~\cite{wav2vec-20} and HuBERT~\cite{hubert-21} extract features from raw waveforms using multiple convolutional layers and then apply MAM or related objectives in the latent space.
As downsampling is applied progressively, severe aliasing-related effects may be less pronounced than in coarse spectrogram patching.
However, because convolutional stacks are composed of learned filters and typically include downsampling operations (e.g., strided convolutions or pooling) without explicit anti-aliasing constraints, their intermediate representations are not guaranteed to be band-limited; as a result, such downsampling can introduce aliasing and reduce shift invariance unless appropriate low-pass filtering is applied~\cite{blurpool-19,anti-aliasing-20}.

\subsection{Temporal Modeling in Audio}

\textbf{State-space Models (SSMs).}\quad S4~\cite{s4-22} and S4D~\cite{s4d-22}, rooted in continuous-time systems theory, unify properties of RNNs and CNNs and are effective for modeling long sequences.
They estimate two linear maps alongside parameters controlling the decay and frequency of damping sinusoids.
Mamba~\cite{mamba-23} extends S4D with a selective mechanism, making those linear maps and certain parameters (including decay and sampling period) input-dependent.
Our work is inspired by the continuous-time signal-analysis perspective underlying S4.
In contrast to S4 and Mamba which target general sequence modeling, our approach aims to analyze aliasing bands in spectrogram subband signals.
Furthermore, we adopt input dependence in the spirit of Mamba; however, in this case, only the decay and frequency are conditioned on the input.

\textbf{Learnable Frontends.}\quad LEAF~\cite{leaf-21}, EfficientLEAF~\cite{efficientleaf-22}, and MuReNN~\cite{murenn-23} replace mel-filterbanks by applying Gabor filters to raw waveforms, with trainable center frequency and window width.
Our approach similarly employs learnable kernels with trainable frequency and window width but operates on spectrogram subbands and is inspired by state-space modeling.
Prior work has reported that frequency parameters in such learnable frontends are sensitive to initialization and exhibit limited drift after training~\cite{notlearn-23}.
Therefore, we hypothesize that vanishing gradients induced by Gaussian windows contribute to this effect, and our analysis motivates adopting a two-sided exponential window to alleviate this gradient-vanishing tendency.

\textbf{Anti-aliasing CNNs.}\quad Applying a low-pass filter before max pooling or strided convolution effectively enhances shift invariance by removing high-frequency content that could cause aliasing upon downsampling~\cite{anti-aliasing-20,improbingsec-21,blurpool-19}.
However, in audio analysis this fixed low-pass strategy can be suboptimal, because high-frequency components often carry salient information (e.g., transients and onsets).
A uniform low-pass filter may suppress useful signal components together with noise.
By contrast, our approach is designed to provide additional features related to high-frequency modulation components that may be underrepresented or distorted after patchification.

\section{Methodology}
In this section, we present AaSP, the proposed aliasing-aware self-supervised pre-training framework.
Its key architectural component is AaPE, which is combined with an EAT-style teacher-student pre-training framework~\cite{eat-24}, a cross-attention predictor, and a contrastive consistency regularization term.
We formulate and evaluate these components as a unified system, because our ablations show that the benefit of AaPE emerges through its interaction with the surrounding learning design rather than from isolated patch-embedding replacement alone.

\subsection{Preliminaries}
{\bf Masked Autoencoder (MAE).}\quad MAE is a self-supervised pre-training framework that learns representations by reconstructing masked portions of an input sequence from the visible tokens.
In the audio domain, Audio-MAE~\cite{audiomae-22} adapts MAE to log-mel spectrograms by patchifying an $F\times T$ spectrogram into non-overlapping $P_\text{freq}\times P_\text{time}$ patches.
Here, $P_\text{freq}$ and $P_\text{time}$ denote the patch sizes along the frequency and time directions, respectively.
Each patch is linearly projected, producing $N_\text{patch}=FT/(P_\text{freq}P_\text{time})$ patch tokens $\mathbf{x}_i \in \mathbb{R}^{D}$ $(i=1,2,\ldots,N_\text{patch})$.
It randomly masks a subset of patches, partitioning the patch set $\mathcal{X}=\{\mathbf{x}_1,\mathbf{x}_2,\ldots,\mathbf{x}_{N_\text{patch}} \}$ into visible and masked subsets $\mathcal{X}_\text{vis}$ and $\mathcal{X}_\text{mask}$ (i.e., $\mathcal{X}_\text{vis} \cup \mathcal{X}_\text{mask}=\mathcal{X}$ and $\mathcal{X}_\text{vis} \cap \mathcal{X}_\text{mask}=\emptyset$).
The encoder consumes the visible patch tokens with positional embeddings and produces a set of output visible tokens $\mathcal{Z}_\text{vis}$.
The predictor forms a length-$N_\text{patch}$ sequence $\{\widetilde{\mathbf{x}}_i\}$ by placing the encoder output at each visible index and a learnable mask token $\mathbf{m}\in \mathbb{R}^D$ at each masked index, followed by adding positional embeddings and reconstructing the original content at masked locations: $\widetilde{\mathbf{x}}_i = \mathbf{z}_i \cdot \mathbb{I}(\mathbf{x}_i \in \mathcal{X}_\text{vis}) + \mathbf{m} \cdot \mathbb{I}(\mathbf{x}_i \in \mathcal{X}_\text{mask})$, where $\mathbf{z}_i$ denotes the encoder output aligned to the original index $i$ when $\mathbf{x}_i$ is visible.
After pre-training, only the encoder is employed for downstream tasks.
Recent work on audio MAE variants often favors self-distillation (teacher-student) over reconstruction~\cite{atst-24,m2d-24,eat-24}, using an EMA teacher and training the encoder (student) to match the teacher’s outputs.

{\bf Structured SSMs.}\quad Structured SSMs define learnable matrices $\mathbf{A} \in \mathbb{R}^{H\times H}$, $\mathbf{B} \in \mathbb{R}^{H\times F}$, and $\mathbf{C} \in \mathbb{R}^{C \times H}$, where $H$ is the hidden-state dimension, $F$ is the input dimension (number of input features), and $C$ is the output dimension.
They model the continuous-time system $\frac{d}{dt}\mathbf{h}(t) = \mathbf{A}\mathbf{h}(t)+\mathbf{B}\mathbf{u}(t)$ and $\mathbf{y}(t) = \mathbf{C}\mathbf{h}(t)$, which maps the input $\mathbf{u}(t) \in \mathbb{R}^F$ at time $t$ to the output $\mathbf{y}(t)\in \mathbb{R}^C$ via the latent state $\mathbf{h}(t)\in \mathbb{R}^H$~\cite{s4-22}.
Applying zero-order-hold (ZOH) discretization and the diagonalization $\mathbf{A} = \mathbf{P}^{-1}\mathbf{\Lambda}\mathbf{P}$ yields the discrete-time system\cite{s4d-22}
\begin{equation}
   \mathbf{y}[n] = \overline{\mathbf{C}}\mathbf{\Lambda}^{-1} \left( e^{\Delta \mathbf{\Lambda}} - \mathbf{I}\right) \sum_{k=0}^n e^{\Delta\mathbf{\Lambda}(k-n)}\overline{\mathbf{B}}\mathbf{u}[k], \label{eq:ssm}
\end{equation}
where $\mathbf{h}(0)=0$, the sampling period is $\Delta \in \mathbb{R}$, the time index is $n=0,1,\ldots,N_\text{time}\!-\!1$, $\overline{\mathbf{B}}=\mathbf{P}^{-1}\mathbf{B}$, $\overline{\mathbf{C}}=\mathbf{C}\mathbf{P}^{-1}$, and $\mathbf{y}[n]=\mathbf{y}(n\Delta)$ denotes the discrete-time output.
The complex diagonal matrix $\mathbf{\Lambda}=\text{diag}(\lambda_1,\ldots,\lambda_H) \in \mathbb{C}^{H \times H}$ has entries $\lambda_i = \alpha_i + j \beta_i$ (with $j$ denoting the imaginary unit), whose real and imaginary parts can be interpreted as a decay $\alpha_i$ and a frequency $\beta_i$ (by multiplying $2\pi$), respectively.
Consequently, the above expression is a causal, one-sided convolution along time with a complex sinusoidal kernel modulated by an exponential window.
For convenience, we write $\alpha,\beta$ in place of $\alpha_i,\beta_i$ when the index is clear.
\begin{figure}
        \centering
        \includegraphics[width=\linewidth]{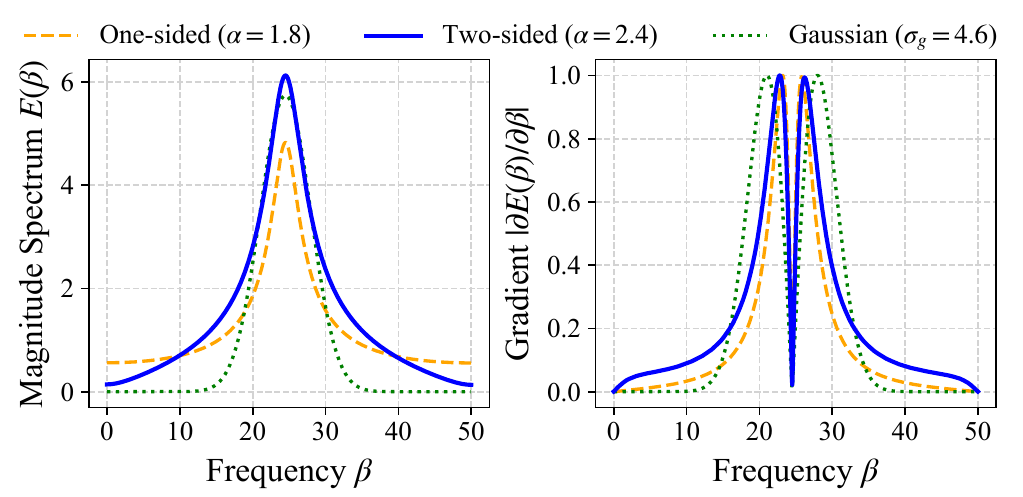}
        \caption{\textbf{Spectra and Gradient Magnitudes for Various Window Functions.} The two-sided window sustains usable gradients over wider frequency offsets, while one-sided and Gaussian windows suffer rapid off-target vanishing; this supports our choice of a two-sided exponential window in SBLU for stable subband-wise estimation in aliasing-prone bands. The window parameters $\alpha$ and $\sigma_g$ are normalized so that the peak gradient magnitude equals 1.0.}
        \label{fig:windows}
\end{figure}
\subsection{Structured Bilateral Laplace Unit (SBLU)} \label{sec:sblu}
{\bf Motivation.} Although aliasing is typically discussed for waveform downsampling, a similar issue arises in practice when patchifying a log-mel spectrogram (mel filterbank spectrogram) along the time axis.
Concretely, after computing an STFT and applying a mel filterbank (and typically a log compression), each mel band forms a discrete-time subband energy sequence sampled at the spectrogram hop $\Delta$ (e.g., 10 ms in our setup).
A strided patch embedding that outputs one token every $S_{\text{time}}$ frames therefore acts as temporal downsampling by a factor of $S_{\text{time}}$ on each subband sequence, reducing the modulation Nyquist limit from $\pi/\Delta$ to $\pi/(\Delta S_{\text{time}})$.
As a result, temporal-variation components in the modulation range $(\pi/(\Delta S_{\text{time}}),\, \pi/\Delta]$ can fold into lower apparent modulation frequencies in the patch-level representation, distorting the subband dynamics presented to the Transformer.
Motivated by this issue, we leverage structured SSMs to derive features from subband signals while targeting the modulation range between the post- and pre-patching Nyquist limits, which is particularly vulnerable under strided patchification.

For stable gradient-based learning of $\beta$ per subband, the gradient with respect to $\beta$ of the SSM-style convolution sum must not vanish.
Here, the convolution sum denotes the kernel-weighted sum obtained by convolving the input-projected subband signal with the complex sinusoidal kernel parameterized by $\beta$ (and decay $\alpha$).
Because this gradient is governed by the window’s frequency response, the one-sided exponential window in Eq.
(\ref{eq:ssm}) suffers from gradient vanishing.
Define a local spectrum at $\beta$ by $E(\beta)=\left| \sum_{k=0}^{N_w-1} w[k]\,x[k]\,e^{-j 2\pi \beta k \Delta} \right|$, for an arbitrary input $x[k]$ and window $w[k]$ of length $N_w$.
For a single-tone input $x[k]=e^{j 2 \pi \omega_0 k \Delta}$, we obtain $E(\beta)=|W(\omega_\delta)|$, where $\omega_\delta=\omega_0-\beta$ and $W(\omega)=\sum_k w[k] e^{j 2 \pi \omega k \Delta}$ is the window’s frequency response.
Thus, $\partial E(\beta)/\partial \beta = -d|W(\omega_\delta)|/d\omega_\delta$, and the gradient magnitude follows the slope of the window’s spectral envelope.
Figure \ref{fig:windows} shows spectra and gradients for a one-sided exponential $w[k]=\exp(-\alpha(N_w-1-k))$\cite{sfe-23}, a two-sided exponential $w[k]=\exp(-\alpha|(N_w-1)/2-k|)$, and a Gaussian $w[k]\propto \exp(-((N_w-1)/2-k)^2/(2\sigma_g^2))$ (as in \cite{leaf-21,murenn-23}) with $\Delta=0.01$, $N_w=127$, and $\omega_0=24.5$.
One-sided exponentials have high average sidelobes owing to edge discontinuity, yet their gradients diminish rapidly away from the target.
Gaussians yield large gradients near the target but very fast sidelobe decay, causing abrupt gradient vanishing.
Two-sided exponentials maintain usable gradient magnitudes across a wider frequency range.

{\bf Formulation.} Based on these observations, we introduce the Structured Bilateral Laplace Unit (SBLU), replacing the one-sided window in Eq. (\ref{eq:ssm}) with a two-sided exponential to reduce gradient vanishing.
Let $\mathbf{u}(t)$ denote the subband signals of a log-mel spectrogram with $F$ mel bands, i.e., the temporal sequences obtained by fixing a mel-frequency band in $\mathbf{X}_\text{mel}\in\mathbb{R}^{F\times T}$ and varying time.
Note that these subband signals are defined on the original spectrogram grid and are conceptually distinct from spectrogram patches/tokens, which aggregate $P_\text{freq}$ adjacent mel bins and $P_\text{time}$ time frames into a single patch.
We then extend Eq.~(\ref{eq:ssm}) to a complex sinusoidal kernel with a two-sided exponential window:
\begin{equation}
        \mathbf{y}[n] = 2\overline{\mathbf{C}} \mathbf{\Lambda}^{-1} \sinh \left( \frac{\Delta\mathbf{\Lambda}}{2}\right)\sum_{k=0}^{K-1} e^{-\Delta\mathbf{\Lambda} \left| k - c \right|} \overline{\mathbf{B}} \mathbf{u}[n+k-c] \label{eq:sblu_complex}
\end{equation}
where the kernel size $K > 1$ is odd and $c=(K-1)/2$ is the center index (derivation in Appendix \ref{sec:appendix_a}).
This operation can be viewed as a discrete bilateral Laplace transform.
The two-sided window sacrifices strict causality but gains locality, thus we implement SBLU as a centered sliding-window convolution.
Because the initial phase depends on the window position, the complex response exhibits phase variability.
We therefore use the response magnitude to suppress phase variation:
\begin{align}
        \mathbf{y}[n] = 2 \overline{\mathbf{C}} \mathbf{\Gamma} \left| \sum_{k=0}^{K-1} e^{-\Delta\mathbf{\Lambda} \left| k - c \right|} \overline{\mathbf{B}} \mathbf{u}[n+k-c] \right|, \label{eq:sblu_abs}\\
        \gamma_i = \sqrt{\frac{\cosh^2(\frac{\Delta\alpha_i}{2}) - \cos^2(\frac{\Delta\beta_i} {2})}{\alpha_i^2 + \beta_i^2}} \label{eq:sblu_gamma}
\end{align}
with $\mathbf{\Gamma}=\text{diag}(\gamma_1,\ldots,\gamma_H)$ (derivation in Appendix \ref{sec:appendix_b}).
As the final output is real-valued, we treat $\overline{\mathbf{B}}$ and $\overline{\mathbf{C}}$ as real trainable matrices, while the convolution itself proceeds in the complex domain.
\begin{figure*}[t]
  \begin{center}
    \includegraphics[width=0.95\textwidth]{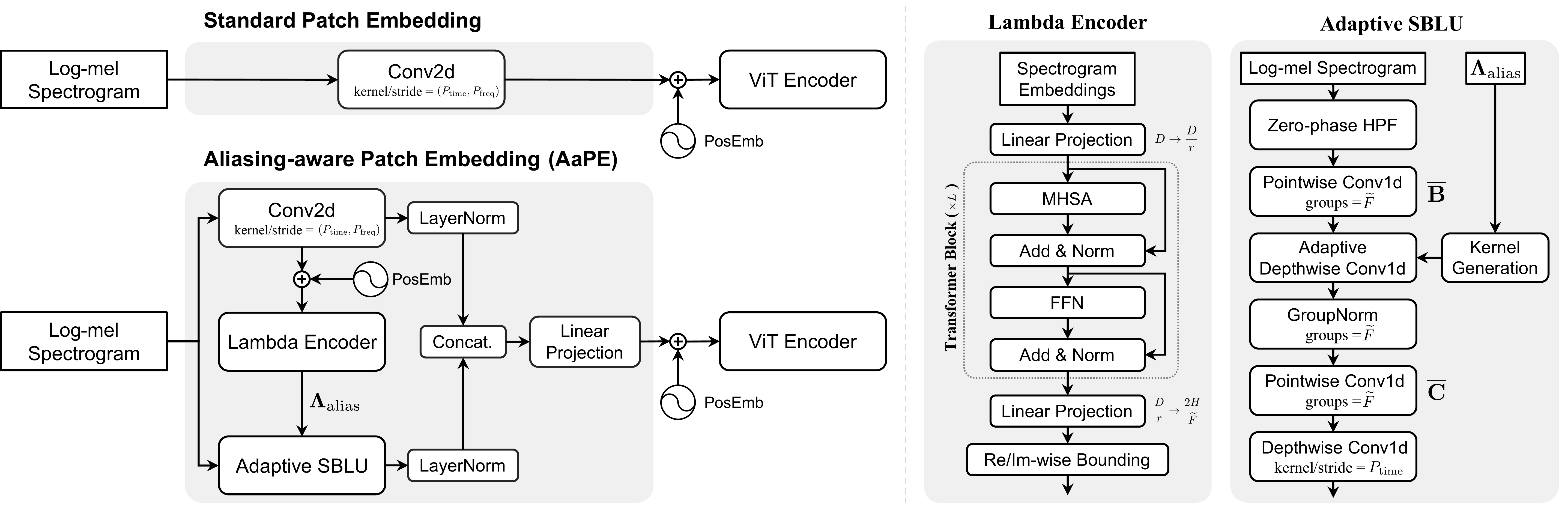}
    \caption{
    \textbf{Architecture of the Aliasing-aware Patch Embedding (AaPE).} AaPE augments the standard ViT patch embedding with aliasing-aware features via three components: (1) Lambda Encoder--takes log-mel spectrogram patches and estimates input-dependent complex kernel parameters $\mathbf{\Lambda}_\text{alias}$ with decay $\alpha$ and frequency $\beta$ per subband using a narrow-and-shallow Transformer and linear projections; (2) Adaptive SBLU--emphasizes aliasing-prone bands with a zero-phase high-pass filter, generates bilateral complex sinusoidal kernels from $\mathbf{\Lambda}_\text{alias}$, and performs memory-efficient adaptive depthwise convolution with grouped pointwise projections to produce an aliasing-focused high-frequency representation; (3) Patch Fusion--independently normalizes and concatenates the standard and aliasing-focused features, then linearly projects to $D$ dimensions to yield the final aliasing-aware patch embedding fed to the ViT encoder.
    }
    \label{fig:sblu}
    \vspace{-18pt}
  \end{center}
\end{figure*}

{\bf Band focusing and stability constraints.} We focus on alias-prone modulation components by restricting the frequency to $\beta \in [\beta_\text{min},\beta_\text{max}]$, where the lower bound is the post-patching Nyquist frequency $\beta_\text{min}=\pi/(\Delta S_{\text{time}})$ and the upper bound is the pre-patching Nyquist frequency $\beta_\text{max}=\pi/\Delta$ (in our non-overlapping setup, $S_{\text{time}}=P_{\text{time}}$).
Thus, the model is designed to analyze the full alias-prone modulation range rather than to preferentially select values near a particular boundary.
Window-edge discontinuity causes ringing in the side lobes of the window’s spectral envelope.
The gradient direction for $\beta$ depends on the local slope of this envelope, and the ringing flips the slope across nearby frequencies.
Consequently, gradient descent on $\beta$ becomes unstable.
Therefore, we impose a positive lower bound on the decay $\alpha$ by requiring edge attenuation $e^{-\Delta \alpha (K-1)/2}\le \epsilon$, yielding $\alpha_\text{min} = -\frac{2\log(\epsilon)}{\Delta (K-1)}$.
We enforce $\alpha \ge \alpha_\text{min}$ and $\beta \in [\beta_\text{min},\beta_\text{max}]$.

\subsection{Aliasing-aware Patch Embedding (AaPE)} \label{sec:aape}
Figure \ref{fig:sblu} illustrates the architecture of AaPE.
We extract high-frequency subband features using SBLU and fuse them with the standard patch tokens to provide additional cues about modulation components that are vulnerable under strided patchification.
To increase expressivity while preserving the structure of patch tokens, we make the learnable decay and frequency parameters in SBLU $\mathbf{\Lambda}$ input-dependent.
This yields a non-stationary kernel that adapts to local amplitude and frequency modulations and other input variations, improving modeling flexibility.
Concretely, our pipeline comprises three components: (1) Lambda Encoder, which adaptively estimates decay and frequency from patch tokens; (2) Adaptive SBLU, which performs frequency analysis conditioned on the estimated parameters; and (3) Patch Fusion, fusing the resulting frequency features with the original patch tokens.

{\bf Lambda Encoder.}\quad We start from the log-mel spectrogram $\mathbf{X}_\text{mel} \in \mathbb{R}^{F \times T}$.
Using non-overlapping patches of size $P_\text{freq}\times P_\text{time}$ (assuming $F$ and $T$ are divisible by $P_\text{freq}$ and $P_\text{time}$), we obtain $\widetilde{F}=F/P_\text{freq}$ frequency patches and $\widetilde{T}=T/P_\text{time}$ temporal patches.
A 2D convolution produces a standard patch embedding $\mathbf{X}_\text{spec} \in \mathbb{R}^{D \times \widetilde{F} \times \widetilde{T}}$.
We add absolute positional embeddings \cite{vit-21} to $\mathbf{X}_\text{spec}$ and estimate input-dependent parameters $\mathbf{\Lambda}_\text{alias} \in \mathbb{C}^{H \times T}$ over the original temporal grid using the Lambda Encoder.
Internally, we first reduce the embedding dimension of $\mathbf{X}_\text{spec}$ from $D$ to $D/r$ via a linear projection, then apply $L$ Transformer blocks; we employ a narrow-and-shallow configuration with $r=6$ and $L=3$ to improve throughput.
The Transformer Block design (attention, number of heads, MLP ratio) is identical to the original ViT architecture~\cite{vit-21}.
Using Transformer Blocks is advantageous in MAE because computations over masked regions can be omitted.
Subsequently, a second linear map projects the output to an embedding dimension of $2H/ \widetilde{F}$, matching the SBLU hidden dimension $H$.
We split the output embeddings along the embedding dimension into two equal parts, interpreted as the real component $\alpha$ and the imaginary component $\beta$ of the complex kernel parameter.
We obtain a lower-bounded decay parameter by applying softplus and a shift: $\alpha_\text{bound} = \text{softplus}(\alpha) + \alpha_\text{min}$.
For the frequency, we bound it to a target interval by applying a sigmoid followed by an affine mapping: $\beta_\text{bound}=(\beta_\text{max}-\beta_\text{min})\sigma(\beta) + \beta_\text{min}$.
We construct a complex-valued parameter tensor by employing the bounded decay as the real part and the bounded frequency as the imaginary part.
To align the parameters with the original temporal resolution while retaining patch-level estimation, we replicate each time-patch entry $P_\text{time}$ times along the temporal axis such that $\widetilde{T}\to T$, yielding a tensor of shape $\mathbb{C}^{\frac{H}{\widetilde{F}}\times \widetilde{F} \times T}$.
Finally, we collapse the frequency-patch axis into the hidden dimension $H$ by reshaping $\mathbb{C}^{\frac{H}{\widetilde{F}}\times \widetilde{F} \times T} \to \mathbb{C}^{H \times T}$ to obtain $\mathbf{\Lambda}_\text{alias}$.

{\bf Adaptive SBLU.}\quad Given the log-mel spectrogram $\mathbf{X}_\text{mel}$ (defined on the original spectrogram time grid) and the input-dependent kernel parameters $\mathbf{\Lambda}_\text{alias}$, this module produces an aliasing-focused representation $\mathbf{X}_\text{alias} \in \mathbb{R}^{\widetilde{C} \times \widetilde{F} \times \widetilde{T}}$.
We use the term aliasing-focused to mean that $\mathbf{X}_\text{alias}$ is designed to encode temporal-modulation components that fall between the post- and pre-patching Nyquist limits, i.e., $(\beta_{\min},\,\beta_{\max}]$, which are prone to folding under strided patchification, rather than the full subband content.
Here, $C$ denotes the output-channel dimension of SBLU (consistent with $\overline{\mathbf{C}}\in\mathbb{R}^{C\times H}$ in Sec.~\ref{sec:sblu}), and we set $\widetilde{C}=C/\widetilde{F}$.
We now describe the role of each operation.

\textit{(a) Band focusing via high-pass filtering:}
to emphasize modulation components in the alias-prone range, we apply a zero-phase high-pass filter along the time axis of each mel-band sequence, implemented by forward--backward filtering with a 63-tap Hamming-window FIR.
The cutoff is set to the post-patching Nyquist limit $\beta_{\min}=\pi/(\Delta S_{\text{time}})$ (Sec.~\ref{sec:sblu}), so that the retained components predominantly capture temporal-modulation (angular) frequencies in $(\beta_{\min},\,\beta_{\max}]$ that cannot be represented after patching and would otherwise fold into lower modulation frequencies.

\textit{(b) Subband mixing while preserving frequency-patch independence:}
we then apply the linear map $\overline{\mathbf{B}}$ to mix mel bins into the SBLU hidden channels.
In practice, $\overline{\mathbf{B}} \in \mathbb{R}^{H\times F}$ and $\overline{\mathbf{C}} \in \mathbb{R}^{C\times H}$ are implemented as grouped $1\times1$ convolutions with $\widetilde{F}$ groups, so that each frequency patch is processed independently and yields its own alias-prone modulation features.

\textit{(c) Localized frequency analysis via adaptive kernels:}
from $\mathbf{\Lambda}_\text{alias}$, we generate complex sinusoidal kernels parameterized by a decay $\alpha$ and a modulation frequency $\beta$ and window them using the two-sided exponential described in Sec.~\ref{sec:sblu}.
We then apply an adaptive depthwise convolution along time (stride $1$ with symmetric zero padding of $(K-1)/2$), yielding a localized, differentiable response magnitude around the selected modulation frequency within the alias-prone band.
Because generating a distinct kernel for every time frame is memory-prohibitive, we generate the kernel on the fly during convolution; we additionally reuse convolution results across parameter-gradient terms via the coupling described in Appendix~\ref{sec:appendix_c}.

\textit{(d) Stabilization and alignment to the patch grid:}
after applying $\overline{\mathbf{C}}$, we use GroupNorm (with $\widetilde{F}$ groups) for training stability while maintaining frequency-patch independence.
Finally, we perform time-only patchification via a depthwise convolution along time (i.e., $\mathbb{R}^{C\times T}\to \mathbb{R}^{C\times \widetilde{T}}$) to align the extracted alias-prone modulation features to the same temporal grid as the standard patch tokens, and reshape $\mathbb{R}^{C \times \widetilde{T}}$ to $\mathbb{R}^{\widetilde{C} \times \widetilde{F} \times \widetilde{T}}$ to obtain $\mathbf{X}_\text{alias}$.
Intuitively, $\mathbf{X}_\text{alias}$ is intended to expose cues about high-modulation dynamics that may be distorted or underrepresented under strided patchification; fusing $\mathbf{X}_\text{alias}$ with $\mathbf{X}_\text{spec}$ makes these cues available to the downstream Transformer without relying on a blanket low-pass filter that may discard task-relevant information.

{\bf Patch Fusion.}\quad After independently normalizing $\mathbf{X}_\text{spec}$ and $\mathbf{X}_\text{alias}$, we concatenate them and apply a linear projection $f_\text{fuse}: \mathbb{R}^{(D+\widetilde{C}) \times \widetilde{F} \times \widetilde{T}}\to \mathbb{R}^{D \times \widetilde{F} \times \widetilde{T}}$ to produce the final aliasing-aware patch tokens $\mathbf{X}_\text{fused} \in \mathbb{R}^{D \times \widetilde{F} \times \widetilde{T}}$, which is fed to the ViT encoder.

\subsection{Predictor and Pre-training Objectives} \label{sec:loss}
As shown in Fig.~\ref{fig:overview}, we learn semantic representations from AaPE and the ViT encoder using a teacher-student framework following EAT~\cite{eat-24}.
The teacher parameters are updated as an EMA of the student, stop-gradient is applied to teacher outputs, and the predictor operates only on the student side.
Our predictor replaces the CNN-based predictor in EAT~\cite{eat-24} with a stack of three cross-attention blocks~\cite{crossmae-24} to facilitate training, each comprising multi-head cross-attention and a position-wise MLP.
During prediction, mask tokens act as queries and attend only to visible tokens (used as keys and values); there is no attention among mask tokens.
This design enables efficient training and demonstrates that cross-attention is effective in the original MAE framework~\cite{crossmae-24} as well as within a teacher-student framework.
We optimize three complementary objectives: a masked-prediction objective, a clip-level alignment objective, and a contrastive regularization term.
We generate $M$ masked views per input sample during pre-training.
The training framework and masking strategy are detailed in Sec.~\ref{sec:training_setup}.

\textbf{Masked Prediction Objective.}\quad The masking loss $\mathcal{L}_m$ encourages reconstruction of latent patch representations by minimizing the L2 distance between the predictor’s outputs on mask tokens and teacher-provided targets.
Consistent with EAT~\cite{eat-24}, the teacher omits a class token.
We use outputs from all teacher layers to form the pooled teacher targets.
We construct the teacher pooled tokens as follows.
In each teacher layer, instance normalization is applied per token and per sample across the embedding dimension, without learnable affine parameters.
These normalized tokens are then averaged across layers, followed by layer normalization across the embedding dimension, also without learnable affine parameters.
The loss $\mathcal{L}_m$ is computed only on the masked tokens, as the cross-attention predictor uses them as queries.

\textbf{Clip-Level Objective.}\quad The clip-level loss $\mathcal{L}_u$ (corresponding to EAT’s utterance-level loss in~\cite{eat-24}) aligns the student’s global representation by minimizing the L2 distance between the encoder’s class token and the teacher’s clip-level target, computed as the mean over the teacher pooled tokens, following~\cite{eat-24}.

\textbf{Contrastive Regularization.}\quad Inspired by SimCLR~\cite{simclr-20}, we introduce a contrastive loss to promote stability and robustness across masked views of the same input sample.
In the context of AaSP, this consistency-promoting signal is intended to help the model use AaPE-derived features more reliably across different masked observations of the same audio.
For each view, the student’s visible tokens are averaged to a single vector, which is fed to a two-layer MLP with hidden dimension equal to half the embedding dimension and GELU activation, followed by a linear output layer; the MLP output is the embedding $\tilde{\mathbf{z}}_i$.
All embeddings are L2-normalized before similarity computation, allowing the dot product to represent cosine similarity.
The temperature is fixed to $\tau=0.2$.
For each $\tilde{\mathbf{z}}_i \in \widetilde{\mathcal{Z}}_\text{vis}$, all pairwise combinations of different masks (views) from the same sample (i.e., ${}_M\mathrm{C}_2$) are treated as positives, and embeddings from other samples in the batch serve as negatives.
The InfoNCE-based contrastive loss~\cite{simclr-20} is
\begin{equation}
        \mathcal{L}_\text{c} = -\frac{1}{N_\text{vis}} \sum_{i=1}^{N_\text{vis}} \frac{1}{|\mathcal{P}(i)|} \sum_{p \in \mathcal{P}(i)} \log \frac{\exp(\tilde{\mathbf{z}}_i^\top \tilde{\mathbf{z}}_p / \tau)}{\sum_{a \in \mathcal{A}(i)} \exp(\tilde{\mathbf{z}}_i^\top \tilde{\mathbf{z}}_a / \tau)},
        \label{eq:loss_contrastive}
\end{equation}
where $N_\text{vis}=|\widetilde{\mathcal{Z}}_\text{vis}|$ is the total number of embeddings across all views in the batch, $\mathcal{P}(i)$ indexes positives for $\tilde{\mathbf{z}}_i$ (same input, different masks), and $\mathcal{A}(i)$ includes $\mathcal{P}(i)$ and all embeddings from other inputs; the anchor $\tilde{\mathbf{z}}_i$ is excluded, and views of the same input that are not in $\mathcal{P}(i)$ are excluded from the denominator.
Dividing by $|\mathcal{P}(i)|$ averages the multi-positive terms for each $\tilde{\mathbf{z}}_i$, making the loss magnitude less sensitive to the number of views.
In terms of computational cost, the dominant term is the similarity evaluation over $\mathcal{A}(i)$, which scales as $O(N_\text{vis}^2)$ with $N_\text{vis}$ (embeddings across the batch and views); the additional summation over positives is $O(N_\text{vis} \cdot |\mathcal{P}(i)|)$, where $|\mathcal{P}(i)|=M-1$ is small in our setup (e.g., $M=4$).
In practice, we compute similarities via a single matrix multiplication of all L2-normalized embeddings.

The final objective is as follows:
\begin{equation}
        \mathcal{L}_\text{total} = \mathcal{L}_m + \eta_u\,\mathcal{L}_u + \eta_c\,\mathcal{L}_c,
\end{equation}
where $\eta_u$ and $\eta_c$ are explicit weighting coefficients for the clip-level alignment and contrastive regularization terms, respectively.
In all experiments, we fix $\eta_u=1$ and tune only $\eta_c$ in the ablations (see Sec.~\ref{sec:ablation}).

\section{Experiments}
\subsection{Datasets} \label{sec:dataset}
Here, we evaluate on a total of ten datasets across multiple evaluation settings: AudioSet in two settings (AS-2M and AS-20K), and nine single-label benchmarks (ESC-50, SCV2, NSynth, US8K, CRM-D, VoxCeleb1, VoxForge, GTZAN, and Surge), covering environmental sound classification, speech command recognition, musical instrument family classification, urban sound classification, speech emotion recognition, speaker identification, language identification, music genre classification, and pitch audio (musical note) classification.
\begin{itemize}
    \item \textbf{AudioSet (AS-2M, AS-20K)}\cite{audioset-17} is a large-scale multi-label dataset of 10-second YouTube clips covering 527 classes. The released training split comprises 2,042,985 unbalanced clips and a 22,176 balanced subset; the evaluation split has 20,383 clips. Owing to missing videos, we retain approximately 1.96M unbalanced, approximately 21k balanced, and approximately 19k evaluation clips. For AS-2M, supervised fine-tuning uses class-balanced sampling (weighted sampling size is 200K). For AS-20K, we fine-tune on the balanced approximately 20k subset (pre-trained on full AS-2M) and report mAP.
    \item \textbf{ESC-50}\cite{esc50-15} is a single-label environmental sound dataset with 2,000 five-second clips across 50 classes. We follow the official 5-fold protocol and report mean accuracy across folds for fine-tuning and linear evaluation.
    \item \textbf{Speech Commands V2 (SCV2)}\cite{spcv2-18} is a single-label keyword spotting dataset comprising 105,829 one-second clips across 35 classes. With the official split (84,843 train, 9,981 validation, 11,005 test), we perform fine-tuning and linear evaluation.
    \item \textbf{NSynth}\cite{nsynth-17} is a single-label instrument family classification dataset with 305,979 four-second clips across 11 classes. With the official split (289,205 train, 12,678 validation, 4,096 test), we perform fine-tuning and linear evaluation.
    \item \textbf{UrbanSound8K (US8K)}\cite{us8k-14} is a single-label urban sound dataset containing 8,732 clips up to four seconds across 10 classes. We conduct linear evaluation with 10-fold cross-validation using the official folds and report mean accuracy.
    \item \textbf{CREMA-D (CRM-D)}\cite{cremad-14} is a single-label emotion recognition dataset with 7,438 utterances of approximately 2.5 s across 6 classes. With the split (5,155 train, 732 validation, 1,551 test) following \cite{m2d-24}, we perform fine-tuning and linear evaluation with the AS-2M-pre-trained model.
    \item \textbf{VoxCeleb1}\cite{voxceleb1-17} is a single-label speaker identification dataset with 153,516 utterances of approximately 8 s across 1,251 classes. With the official split (138,361 train, 6,904 validation, 8,251 test) following \cite{m2d-24}, we perform linear evaluation.
    \item \textbf{VoxForge}\cite{voxforge-18} is a single-label language identification dataset with 176,428 approximately 6 s utterances of varying lengths across 6 classes. With the split (121,281 train, 26,684 validation, 28,463 test) following \cite{m2d-24}, we perform linear evaluation.
    \item \textbf{GTZAN}\cite{gtzan-02} is a single-label music genre classification dataset with 930 30-second clips across 10 classes. With the split (443 train, 197 validation, 290 test) following \cite{m2d-24}, we perform linear evaluation.
    \item \textbf{Surge}\cite{surge-21} is a single-label pitch audio classification dataset (Surge synthesizer) with 183,392 four-second clips across 88 classes (88 MIDI notes). With the split (148,896 train, 17,160 validation, 17,336 test) following \cite{m2d-24}, we perform fine-tuning and linear evaluation.
\end{itemize}
\setlength{\tabcolsep}{2.5pt}
\begin{table*}[t]
        \caption{Hyperparameters for Pre-training and Fine-tuning.}
\centering
\begin{threeparttable}
\begin{tabular}{l|c|ccccccc}
\toprule
\multirow{2}{*}{Hyperparameters} & Pre-training & \multicolumn{7}{c}{Fine-tuning}  \\
 & AS-2M & AS-2M & AS-20K & ESC-50 & SCV2 & CRM-D & NSynth & Surge \\
\midrule
Optimizer & \multicolumn{8}{c}{AdamW~\cite{adamw-17} with $\beta_1 = 0.9, \beta_2 = 0.95$} \\
LR schedule & \multicolumn{8}{c}{Cosine Annealing~\cite{cossched-16} with linear warm-up} \\
Peak LR & $5.0\times10^{-4}$ & \multicolumn{7}{c}{$5.0\times10^{-5}$} \\
Minimum LR & 0.0 & $1.0\times10^{-6}$ & $1.0\times10^{-6}$ & $1.0\times10^{-7}$ & $1.0\times10^{-7}$ & $1.0\times10^{-7}$ & $1.0\times10^{-7}$ & $1.0\times10^{-7}$ \\
Layer decay~\cite{beit-21} & N/A & 0.70 & 0.70 & 0.65 & 0.75 & 0.65 & 0.65 & 0.65 \\
Weight decay & 0.05 & 0.05 & 0.05 & 0.05 & 0.05 & 0.05 & 0.05 & 0.05 \\
Training steps & 600K & 300K & 40K & 8K & 160K & 8K & 443K & 233K\\
Warm-up steps & 80K & 30K & 4K & 800 & 16K & 800 & 44K & 23K \\
Batch size\tnote{1} & 512 & 128 & 128 & 128 & 128 & 128 & 128 & 128 \\
\# of GPUs & 4 & 4 & 4 & 1 & 1 & 1 & 1 & 1 \\
Mixup~\cite{mixup-18} & 0.0 & 0.8 & 0.8 & 0.0 & 0.8 & 0.0 & 0.0 & 0.0 \\
Drop path~\cite{droppath-16} & 0.0 & 0.1 & 0.1 & 0.1 & 0.1 & 0.1 & 0.1 & 0.1 \\
\bottomrule
\end{tabular}
\begin{tablenotes}
        \footnotesize
        \item[1] Batch size means the total batch size for the multi-GPU training.
\end{tablenotes}
\end{threeparttable}
\label{tab:hyperparams_1}
\end{table*}
\setlength{\tabcolsep}{2.5pt}
\begin{table*}[t]
        \caption{Hyperparameters for Linear Evaluation.}
\centering
\begin{tabular}{l|ccccccccc}
\toprule
\multirow{2}{*}{Hyperparameters}  & \multicolumn{9}{c}{Linear Evaluation}  \\
 & ESC-50 & US8K & SCV2 & VoxCeleb1 & VoxForge & CRM-D & GTZAN & NSynth & Surge \\
\midrule
Optimizer & \multicolumn{9}{c}{Adam with $\beta_1 = 0.9, \beta_2 = 0.999$} \\
LR schedule & \multicolumn{9}{c}{Cosine Annealing\cite{cossched-16}} \\
Peak LR & $5.0\times10^{-4}$ & $1.0\times10^{-4}$ & $1.0\times10^{-4}$ & $1.0\times10^{-4}$ & $1.0\times10^{-4}$ & $1.0\times10^{-4}$ & $5.0\times10^{-5}$ & $1.0\times10^{-4}$ & $1.0\times10^{-4}$ \\
Minimum LR & 0.0 & 0.0 & 0.0 & 0.0 & 0.0 & 0.0 & 0.0 & 0.0 & 0.0 \\
Weight decay & 0.0 & 0.0 & 0.0 & 0.0 & 0.0 & 0.0 & 0.0 & 0.0 & 0.0 \\
Training steps & 3K & 12K & 133K & 216K & 190K & 8K & 6K & 443K & 233K \\
Batch size & 128 & 128 & 128 & 128 & 128 & 128 & 16 & 128 & 128 \\
\# of GPUs & 1 & 1 & 1 & 1 & 1 & 1 & 1 & 1 & 1 \\
\bottomrule
\end{tabular}
\label{tab:hyperparams_2}
\end{table*}
\subsection{Training Setup} \label{sec:training_setup}
We summarize the protocols for pre-training, fine-tuning, and linear evaluation below.
The hyperparameters used for pre-training and fine-tuning are listed in Table~\ref{tab:hyperparams_1}, while those for linear evaluation are shown in Table~\ref{tab:hyperparams_2}.
All fine-tuning or linear evaluation experiments use a model pre-trained on AS-2M without label information.
Unless otherwise noted, all audio is resampled to 16 kHz, and log-mel spectrograms are computed with a 400-sample Hann window, 160-sample hop, 128 mel-bins ($F=128$), and an FFT size of 1,024.
For SBLU, we set the kernel size $K=63$, the step size $\Delta=0.01$, the patch size $P_\text{time}=P_\text{freq}=16$, the SBLU hidden dimension $H=128$, the output dimension $C=1,024$ and the window attenuation threshold $\epsilon=0.01$.
The contrastive regularization loss weight $\eta_c$ is set to $0.1$ by default.
Log-mel spectrograms are standardized using mean $-7.1$ and standard deviation $4.1$.
Training is conducted on NVIDIA H100-80GB GPUs.

\textbf{Pre-training Setup.}\quad In AaSP, we employ a teacher-student framework to learn representations from the acoustic patch sequence produced by AaPE.
The student encoder is a 12-layer Vision Transformer \cite{vit-21}.
The predictor consists of the 3 cross-attention blocks as in CrossMAE~\cite{crossmae-24}, with hidden size 512 and 16 heads.
Queries are randomly selected from masked tokens; we use 75\% of the masked tokens as queries to facilitate training as in \cite{crossmae-24}.
We apply inverse block masking \cite{eat-24} with an 80\% token mask ratio and a multi-mask scheme with $M=4$ masks (views) per input; all $M$ views are included in the same batch.
For each view, we sample a block size uniformly from $\{3\times8,~4\times6,~5\times5\}$ in (frequency $\times$ time) patches; the remaining tokens required to reach the 80\% mask ratio are selected via non-block random masking.
Block locations and the residual random masks are sampled independently for each of the $M$ views.
The teacher branch is stop-gradient and updated via an EMA of the student using a cosine momentum schedule identical to \cite{dinov2-23}.
Specifically, $\theta_\text{teacher}^{(m)} \leftarrow \xi^{(m)}\,\theta_\text{teacher}^{(m-1)} + (1-\xi^{(m)})\,\theta_\text{student}^{(m)}$ where $m$ is the iteration step and $\xi^{(m)}$ increases from 0.994 to 1.0 over the maximum iterations as in \cite{dinov2-23}.
Teacher signals (pooled tokens and their mean-pooled utterance target) are formed using functional instance and layer normalization without learnable affine parameters.
AS-2M clips are 10 s; for efficiency, we randomly crop 6-second segments and set the log-mel spectrogram length to 608 time frames by truncation or zero-padding, following prior work \cite{atst-24, m2d-24}.
We utilize 2D absolute positional embedding (time-frequency) as in \cite{audiomae-22}.
For downstream inputs shorter than the pre-training length, embeddings are truncated; for longer inputs, they are extended via linear interpolation along the time axis, leaving the frequency axis unchanged.

\textbf{Fine-tuning Setup.}\quad Following \cite{eat-24}, we append a LayerNorm and a linear classification head on the class token, optimizing all parameters end-to-end with a small learning rate.
We adopt layer-wise learning-rate decay (Layer decay\cite{beit-21}) as in \cite{eat-24,atst-24}, assigning smaller learning rates to earlier layers.
For data augmentation, Mixup\cite{mixup-18} and SpecAugment\cite{specaug-19} are applied, with the mask size of one-fifth the input spectrogram size; no data augmentation is used for ESC-50.
During fine-tuning except for ESC-50, an EMA of model weights with decay 0.995 is maintained for stable inference.
Loss functions follow the task: multi-label datasets use BCE-with-logits; single-label datasets use cross-entropy unless Mixup is applied, in which case we use BCE.

\textbf{Linear Evaluation Setup.}\quad In this setup, we freeze the pre-trained encoder and only train a linear classifier on top of the frozen features.
We conduct linear evaluation using the EVAR platform\footnote{https://github.com/nttcslab/eval-audio-repr} and follow its protocol unless otherwise stated.
For the patch-token representation, we use the same aggregation method as in M2D and MATPAC.
We average the patch tokens along the time axis and then concatenate them across the frequency axis.
Additionally, in our model, we use the class token learned during pre-training.
We concatenate the class token with the aggregated patch-token representation to form the final feature vector for the linear classifier.
As in \cite{atst-24}, no data augmentation is used for linear evaluation.
The features obtained from the pre-trained encoder are collectively standardized, immediately prior to being input to the linear classifier, using pre-computed statistics over the entire training set.

All downstream evaluation results (fine-tuning, linear evaluation, and ablations) are averaged over three random seeds and reported as mean $\pm$ standard deviation.
For all evaluation scores, we report the best result achieved within the specified number of training steps.

\subsection{Evaluation Results} \label{sec:result}
We evaluate the full AaSP framework against representative strong self-supervised baselines with broadly comparable spectrogram-based settings and model scale. We do not reproduce every related EMA-based audio SSL method, as a fully unified comparison would require substantial re-alignment of training and evaluation protocols and is beyond the scope of this work.

\setlength{\tabcolsep}{5pt}
\begin{table*}[t]
        \caption{Performance Comparison on Fine-tuning.}
\centering
\begin{tabular}{lccccccccc}
\toprule
\multirow{2}{*}{Model} & \multirow{2}{*}{\#Params.} & AS-2M & AS-20K & ESC-50 & SCV2 & CRM-D & NSynth & Surge \\
& & mAP(\%) & mAP(\%)  & Acc(\%)   & Acc(\%) & Acc(\%) & Acc(\%) & Acc(\%) \\
\midrule
MaskSpec\cite{maskspec-23}& 86M & 47.1 & 32.3 & 89.6  & 97.7 & - & - & - \\
data2vec \cite{data2vec-22}& 94M & - & 34.5 & -  & - & - & - & -\\
Audio-MAE \cite{audiomae-22}& 86M & 47.3 & 37.1 & 94.1 & 98.3 & - & - & - \\
M2D/0.7\cite{m2d-24} & 86M & 47.9 & 38.6 & 96.0 & 98.4 & - & - & - \\
ASiT\cite{asit-24}& 86M & 48.0 & 38.3 & 95.6 & \textbf{98.9} & - & - & - \\
BEATs${}_{iter3}$\cite{beats-23} & 90M & 48.0 & 38.3 & 95.6 & 98.3 & - & - & - \\
ATST-Frame\cite{atst-24} & 86M & 48.0 & 39.0 & - & 98.1 & - & 79.2 & - \\
EAT\cite{eat-24} & 88M & 48.6 & 40.2 & 95.9 & 98.3 & \textbf{74.9}${\scriptstyle \pm 0.4}$ & - & \textbf{53.2}${\scriptstyle \pm 0.6}$ \\
ASDA\cite{asda-25} & 93M & 49.0 & 41.5 & 96.1 & 98.3 & - & - & - \\
SSLAM\cite{sslam-25} & 88M & \textbf{50.2} & 40.9 & 96.2 & 98.1 & - & - & - \\
\midrule
AaSP (ours) & 87M & 49.8${\scriptstyle \pm 0.0}$ & \textbf{41.9}${\scriptstyle \pm 0.1}$ & \textbf{97.5}${\scriptstyle \pm 0.1}$ & 97.9${\scriptstyle \pm 0.0}$ & 73.4${\scriptstyle \pm 0.2}$ & \textbf{88.7}${\scriptstyle \pm 0.1}$ & 52.3${\scriptstyle \pm 0.1}$ \\
\bottomrule
\end{tabular}
\label{tab:exp_finetuning}
\end{table*}
\textbf{Performance for Fine-tuning.}\quad Table~\ref{tab:exp_finetuning} summarizes fine-tuning results of self-supervised models pre-trained on AudioSet across downstream tasks spanning multiple audio domains.
On general acoustic/environmental sound benchmarks (AS-2M, AS-20K, and ESC-50), AaSP shows clear gains over closely related architecture- or training-design-oriented baselines, improving over EAT on AS-2M (49.8\% vs. 48.6\%) and AS-20K (41.9\% vs. 40.2\%), and outperforming ASDA on both AS-2M (49.8\% vs. 49.0\%) and AS-20K (41.9\% vs. 41.5\%). On ESC-50, AaSP achieves the best result in Table~\ref{tab:exp_finetuning}. SSLAM is retained as a reference point, but its mixture-based objective is orthogonal to our contribution.
We note that SSLAM’s gains typically derive from its mixture-based pre-training objective, whereas our contribution lies in the full aliasing-aware pre-training framework, within which the proposed patch-embedding design serves as one key architectural component.
These approaches are orthogonal and non-exclusive; combining AaSP with SSLAM is therefore a promising direction for future work.
On speech-related tasks (SCV2 and CRM-D), AaSP remains competitive but shows smaller gains: on SCV2 it reaches 97.9\% accuracy, slightly below SSLAM (98.1\%) and ASiT (98.9\%; $-1.0$ percentage points), and on CRM-D it achieves 73.4\% accuracy compared with EAT’s 74.9\%.
We hypothesize that, because SCV2 consists of 1-second signals, AaSP may not fully leverage frequency cues within such short contexts, and that speech-specific cues may further limit the benefit of aliasing-focused features.
On music-related tasks (NSynth and Surge), AaSP yields clear improvements on NSynth (88.7\% accuracy; substantially higher than ATST-Frame’s 79.2\%), consistent with the hypothesis that frequency-sensitive representations may particularly benefit periodic or quasi-stationary musical signals, while it remains competitive on Surge (52.3\% accuracy, close to EAT’s 53.2\%).
\setlength{\tabcolsep}{5pt}
\begin{table*}[t]
        \caption{Performance Comparison on Linear Evaluation.}
\centering
\begin{tabular}{lcccccccccccc}
\toprule
\multirow{2}{*}{Model} & \multirow{2}{*}{\#Params.} & ESC-50 & US8K & SCV2 & VoxCeleb1 & VoxForge & CRM-D & GTZAN & NSynth & Surge \\
&  & Acc(\%) & Acc(\%) & Acc(\%) & Acc(\%) & Acc(\%) & Acc(\%) & Acc(\%) & Acc(\%) & Acc(\%) \\
\midrule
BEATs${}_{iter3}$\cite{beats-23} & 90M &
86.9${\scriptstyle \pm 1.4}$ & 84.8${\scriptstyle \pm 0.1}$ & 89.4${\scriptstyle \pm 0.1}$ &
41.4${\scriptstyle \pm 0.7}$ & 94.1${\scriptstyle \pm 0.3}$ & 64.7${\scriptstyle \pm 0.8}$ &
72.6${\scriptstyle \pm 4.3}$ & 75.9${\scriptstyle \pm 0.2}$ & 39.3${\scriptstyle \pm 0.4}$ \\
ATST-Frame\cite{atst-24} & 86M &
90.9${\scriptstyle \pm 0.6}$ & 85.8                         & 94.9                         &
\textbf{77.4}                & \textbf{98.8}${\scriptstyle \pm 0.3}$ & 72.3${\scriptstyle \pm 0.7}$ &
82.9${\scriptstyle \pm 6.0}$ & 75.9${\scriptstyle \pm 0.0}$ & 40.6${\scriptstyle \pm 0.2}$ \\
M2D/0.7\cite{m2d-24} & 86M &
91.3${\scriptstyle \pm 0.6}$ & 87.6${\scriptstyle \pm 0.2}$ & \textbf{96.0}${\scriptstyle \pm 0.1}$ &
73.4${\scriptstyle \pm 0.2}$ & 98.3${\scriptstyle \pm 0.0}$ & \textbf{73.0}${\scriptstyle \pm 0.7}$ &
84.1${\scriptstyle \pm 2.7}$ & 75.7${\scriptstyle \pm 0.1}$ & 42.1${\scriptstyle \pm 0.2}$ \\
EAT\cite{eat-24} & 88M &
85.6${\scriptstyle \pm 0.4}$ & 81.7${\scriptstyle \pm 0.3}$ & 81.5${\scriptstyle \pm 0.4}$ &
39.6${\scriptstyle \pm 0.5}$ & 92.6${\scriptstyle \pm 0.0}$ & 64.9${\scriptstyle \pm 2.8}$ &
73.7${\scriptstyle \pm 0.5}$ & 71.9${\scriptstyle \pm 0.1}$ & 39.0${\scriptstyle \pm 0.3}$ \\
SSLAM\cite{sslam-25} & 88M &
85.7${\scriptstyle \pm 1.3}$ & 81.4${\scriptstyle \pm 0.3}$ & 82.5${\scriptstyle \pm 0.4}$ &
33.9${\scriptstyle \pm 0.8}$ & 92.0${\scriptstyle \pm 0.1}$ & 66.4${\scriptstyle \pm 1.5}$ &
70.9${\scriptstyle \pm 2.5}$ & 73.8${\scriptstyle \pm 0.7}$ & 38.0${\scriptstyle \pm 0.9}$ \\
MATPAC/n10\cite{matpac-25} & 86M &
\textbf{93.5}${\scriptstyle \pm 0.1}$ & 89.4${\scriptstyle \pm 0.1}$ & - & - & - & - &
\textbf{85.3}${\scriptstyle \pm 0.4}$ & 74.3${\scriptstyle \pm 0.2}$ & - \\
\midrule
AaSP (ours) & 87M &
88.2${\scriptstyle \pm 0.6}$ & \textbf{89.9}${\scriptstyle \pm 0.1}$ & 89.2${\scriptstyle \pm 0.1}$ &
49.6${\scriptstyle \pm 0.1}$ & 88.9${\scriptstyle \pm 0.0}$ & 68.5${\scriptstyle \pm 0.4}$ &
82.2${\scriptstyle \pm 0.0}$ & \textbf{79.4}${\scriptstyle \pm 0.1}$ & \textbf{42.6}${\scriptstyle \pm 0.1}$ \\
\bottomrule
\end{tabular}
\label{tab:exp_linear_evaluation}
\end{table*}

\textbf{Performance for Linear Evaluation.}\quad Table~\ref{tab:exp_linear_evaluation} summarizes linear evaluation results on nine benchmarks, grouped into environmental sound (ESC-50, US8K), speech (SCV2, VoxCeleb1, VoxForge, CRM-D), and music (GTZAN, NSynth, Surge).
On the environmental sound benchmarks, AaSP achieves strong performance, reaching 89.9\% on US8K, which exceeds MATPAC and M2D, while remaining competitive on ESC-50.
On the speech benchmarks, AaSP is generally less competitive than the strongest baselines (e.g., 49.6\% on VoxCeleb1 and 88.9\% on VoxForge) and remains below M2D on CRM-D, suggesting that the aliasing-focused features introduced by SBLU do not consistently translate into gains for speech-related representations under this linear protocol.
On the music benchmarks, AaSP shows clear advantages on periodic or quasi-stationary signals, most notably achieving 79.4\% on NSynth and outperforming ATST-Frame and BEATs${}_{iter3}$, which is consistent with the hypothesis that frequency-sensitive feature extraction can be beneficial for such musical content; AaSP is also competitive on GTZAN and improves over several teacher-student masked-modeling baselines on Surge.
In addition, when compared within the EAT-family methods that share the clip-level objective (e.g., EAT and SSLAM), AaSP remains competitive and improves performance on several benchmarks, suggesting that the observed gains are not explained by the clip-level objective alone and are more consistent with the full pre-training design.

Overall, AaSP shows competitive transfer under both fine-tuning and linear evaluation across diverse domains, with clearer gains on general acoustic and music benchmarks where periodic structure and high-frequency/modulation cues are salient, while improvements on speech-related benchmarks are smaller and sometimes absent under the same protocols.

\subsection{Ablation Studies} \label{sec:ablation}
We assess how each major component of the proposed method contributes to the final performance.

\setlength{\tabcolsep}{3.0pt}
\begin{table}[t]
        \caption{Ablation of Patch Embedding and Contrastive Regularization for Fine-tuning.}
\centering
\begin{threeparttable}
\begin{tabular}{cc|ccccc}
\toprule
Patch & Weight & \multirow{2}{*}{\#Params.\tnote{2}} & AS-2M & AS-20K & ESC-50 & CRM-D\\
Embedding\tnote{1} & $\eta_c$ & & mAP(\%) & mAP(\%) & Acc(\%) & Acc(\%)\\
\midrule
SPE & 0.1  & 0.43M & 45.8${\scriptstyle \pm 0.1}$ & 35.6${\scriptstyle \pm 0.1}$ & 93.6${\scriptstyle \pm 0.3}$ & 68.3${\scriptstyle \pm 0.8}$ \\
SPE & 0.01 & 0.43M & 49.4${\scriptstyle \pm 0.1}$ & 41.0${\scriptstyle \pm 0.0}$ & 96.4${\scriptstyle \pm 0.3}$ & 72.4${\scriptstyle \pm 0.5}$ \\
SPE & 0.001 & 0.43M & 48.6${\scriptstyle \pm 0.0}$ & 38.6${\scriptstyle \pm 0.3}$ & 95.1${\scriptstyle \pm 0.1}$ & 71.4${\scriptstyle \pm 0.3}$ \\
SPE & 0  & 0.43M & 48.0${\scriptstyle \pm 0.1}$ & 37.6${\scriptstyle \pm 0.1}$ & 93.4${\scriptstyle \pm 0.1}$ & 68.4${\scriptstyle \pm 0.1}$ \\
\rowcolor{gray!15}
AaPE\tnote{3} & 0.1 & 1.86M & \textbf{49.8}${\scriptstyle \pm 0.0}$ & \textbf{41.9}${\scriptstyle \pm 0.1}$ & \textbf{97.5}${\scriptstyle \pm 0.1}$ & 73.4${\scriptstyle \pm 0.2}$ \\
AaPE & 0.01 & 1.86M & 49.7${\scriptstyle \pm 0.1}$ & 41.7${\scriptstyle \pm 0.1}$ & 97.3${\scriptstyle \pm 0.2}$ & \textbf{73.8}${\scriptstyle \pm 0.4}$ \\
AaPE & 0.001 & 1.86M & 48.3${\scriptstyle \pm 0.2}$ & 38.7${\scriptstyle \pm 0.0}$ & 96.0${\scriptstyle \pm 0.1}$ & 70.5${\scriptstyle \pm 0.2}$ \\
AaPE & 0 & 1.86M & 48.0${\scriptstyle \pm 0.0}$ & 37.0${\scriptstyle \pm 0.2}$ & 94.5${\scriptstyle \pm 0.2}$ & 69.2${\scriptstyle \pm 0.5}$ \\
\bottomrule
\end{tabular}
\begin{tablenotes}
        \footnotesize
        \item[1] All rows use the same AaSP learning design; only the patch embedding (SPE vs.\ AaPE) and the contrastive regularization weight $\eta_c$ are varied.
        \item[2] Parameter counts reflect patch-embedding layer parameters only.
        \item[3] Gray-shaded row denotes the default configuration.
\end{tablenotes}
\end{threeparttable}
\label{tab:exp_ablation_pe_loss}
\end{table}
        \setlength{\tabcolsep}{6.0pt}
\begin{table}[t]
        \caption{Ablation of Patch Embedding Replacement in EAT for Fine-tuning.}
\centering
\begin{threeparttable}
\begin{tabular}{c|cccc}
\toprule
\multirow{2}{*}{Model Setting} & AS-2M & AS-20K & ESC-50 & CRM-D \\
& mAP(\%) & mAP(\%) & Acc(\%) & Acc(\%) \\
\midrule
EAT {\footnotesize (our impl.)} & \textbf{49.0}${\scriptstyle \pm 0.1}$ & \textbf{38.5}${\scriptstyle \pm 0.2}$ & \textbf{95.5}${\scriptstyle \pm 0.2}$ & \textbf{74.9}${\scriptstyle \pm 0.4}$ \\
EAT with AaPE\tnote{1} & 47.5${\scriptstyle \pm 0.1}$ & 36.4${\scriptstyle \pm 0.1}$ & 92.9${\scriptstyle \pm 0.3}$ & 71.1${\scriptstyle \pm 1.0}$ \\
\bottomrule
\end{tabular}
\begin{tablenotes}
        \footnotesize
        \item[1] In this setting, AaPE replaces the standard patch embedding in EAT. All other components follow the original design; we do not use contrastive regularization or the cross-attention predictor.
\end{tablenotes}
\end{threeparttable}
\label{tab:exp_ablation_patch_embedding_replacement}
\end{table}
\textbf{Patch Embedding and Contrastive Weight.}\quad Table~\ref{tab:exp_ablation_pe_loss} shows that the shared AaSP learning design already yields strong downstream performance with the standard patch embedding (SPE), and that replacing SPE with our Aliasing-aware Patch Embedding (AaPE) further improves the results.
The table provides a controlled comparison, since the SPE and AaPE rows use the same predictor, masking strategy, and objectives, differing only in the patch embedding, with contrastive regularization weights $\eta_c \in \{0.1,~0.01,~0.001,~0\}$.
Under this matched setting, AaPE does not consistently outperform SPE when the contrastive term is disabled ($\eta_c=0$), suggesting that the benefit of AaPE becomes more pronounced when contrastive regularization is active.
With SPE, $\eta_c=0.01$ gives the best overall performance, whereas increasing it to $\eta_c=0.1$ degrades all metrics, indicating that SPE is sensitive to overly large contrastive weights.
In contrast, AaPE remains robust at $\eta_c=0.1$ and improves performance as $\eta_c$ increases on AS-2M, AS-20K, and ESC-50; on CRM-D, the best result is obtained at $\eta_c=0.01$, with only a slight drop at $\eta_c=0.1$.
Comparing the best setting for each embedding, AaPE ($\eta_c=0.1$) consistently outperforms SPE ($\eta_c=0.01$) across all benchmarks, improving from 49.4 to 49.8 mAP on AS-2M, from 41.0 to 41.9 mAP on AS-20K, from 96.4\% to 97.5\% on ESC-50, and from 72.4\% to 73.8\% on CRM-D.
We also note the computational overhead of AaPE: under our default input and patch configuration ($F=128$, $T=608$, $P_\text{freq}=P_\text{time}=16$, $D=768$, $K=63$), SPE (a single strided Conv2D) costs approximately 0.12 GFLOPs per clip for a forward pass, whereas AaPE costs approximately 0.55 GFLOPs.\footnote{See Appendix~\ref{sec:appendix_flops} for the FLOPs estimation details (reported for the masked-input pre-training regime).} This additional cost is confined to the patch-embedding stage and does not increase the Transformer encoder complexity; moreover, during pre-training the Lambda Encoder is applied only to visible tokens, reducing its cost with higher mask ratios.

Complementary to Table~\ref{tab:exp_ablation_pe_loss}, which compares SPE and AaPE under the same AaSP learning design, Table~\ref{tab:exp_ablation_patch_embedding_replacement} isolates the effect of replacing only the patch embedding inside the original EAT framework. Specifically, it reports an otherwise identical EAT configuration where we replace the standard patch embedding with AaPE (“EAT with AaPE”), while keeping the original EAT predictor and objectives (i.e., without the cross-attention predictor or contrastive regularization).
In this setting, replacing the patch embedding alone consistently degrades performance across all four benchmarks, even though the rest of the training pipeline remains the original EAT design.
This result shows that AaPE should not be interpreted as a universal drop-in replacement for the original EAT patch embedding.
At the same time, Table~\ref{tab:exp_ablation_pe_loss} shows that under the same AaSP learning design, replacing SPE with AaPE improves performance on several benchmarks.
Taken together, these results suggest that the gains of AaSP are not explained by patch-embedding replacement alone, but are more consistent with an interaction between AaPE and the modified learning design of AaSP, including the cross-attention predictor, multi-mask training, and contrastive regularization.
In particular, our results are consistent with the view that consistency-promoting learning signals may help the model make more effective use of AaPE-derived features, although the present experiments do not fully isolate the contribution of each component.

\setlength{\tabcolsep}{2.25pt}
\begin{table}[t]
        \caption{Ablation of SBLU Adaptivity for Fine-tuning.}
\centering
\begin{threeparttable}
\begin{tabular}{c|cccccc}
\toprule
SBLU & \multirow{2}{*}{\#Params.\tnote{3}} & AS-2M & AS-20K & ESC-50 & CRM-D & Surge \\
Adaptivity &  & mAP(\%) & mAP(\%) & Acc(\%) & Acc(\%) & Acc(\%) \\
\midrule
Non-adaptive\tnote{1} & 1.16M & 49.7${\scriptstyle \pm 0.0}$ & 41.7${\scriptstyle \pm 0.1}$ & 97.2${\scriptstyle \pm 0.0}$ & 73.0${\scriptstyle \pm 0.5}$ & \textbf{52.6}${\scriptstyle \pm 0.1}$ \\
\rowcolor{gray!15}
Adaptive\tnote{2} & 1.86M & \textbf{49.8}${\scriptstyle \pm 0.0}$ & \textbf{41.9}${\scriptstyle \pm 0.1}$ & \textbf{97.5}${\scriptstyle \pm 0.1}$ & \textbf{73.4}${\scriptstyle \pm 0.2}$ & 52.3${\scriptstyle \pm 0.1}$ \\
\bottomrule
\end{tabular}
\begin{tablenotes}
        \footnotesize
        \item[1] In the non-adaptive setting, the Lambda encoder is disabled and its frequency/decay parameters are learned as static, input-independent values.
        \item[2] Gray-shaded row denotes the default configuration.
        \item[3] Parameter counts reflect patch-embedding layer parameters only.
\end{tablenotes}
\end{threeparttable}
\label{tab:exp_ablation_adaptivity}
\end{table}
\textbf{Adaptivity of SBLU.}\quad Within AaPE, SBLU extracts aliasing-focused frequency features from log-mel subband time series and then aligns them to the patch grid (Sec.~\ref{sec:aape}).
We introduce Adaptive SBLU, which estimates the kernel parameters (decay and frequency) for each time patch via the Lambda encoder.
In the non-adaptive ablation, these parameters are replaced with learnable values that are fixed across time patches.
Table~\ref{tab:exp_ablation_adaptivity} reports results on both diverse audio benchmarks (AS-2M, AS-20K, and ESC-50) and more domain-biased single-label tasks, i.e., speech emotion recognition (CRM-D) and pitch audio classification (Surge).
On four of the five tasks (AS-2M, AS-20K, ESC-50, and CRM-D), adaptivity provides modest gains (e.g., +0.1 mAP on AS-2M, +0.2 mAP on AS-20K, +0.3 Acc on ESC-50, and +0.4 Acc on CRM-D).
In contrast, on Surge the non-adaptive setting is slightly better (+0.3 Acc).
These results suggest that input-dependent decay/frequency estimation may be more helpful when temporal modulation patterns vary substantially across inputs, while for more stationary signals a fixed parameterization may sometimes be sufficient.
Overall, adaptivity provides modest improvements on most tasks at the cost of higher module complexity (1.86M vs 1.16M parameters in the patch embedding); when model size or latency is constrained, non-adaptive SBLU remains a competitive simplified option.

\setlength{\tabcolsep}{7pt}
\begin{table}[t]
        \caption{Ablation of SBLU Kernel Size for Fine-tuning.  }
\centering
\begin{threeparttable}
\begin{tabular}{l|cccc}
\toprule
Kernel size & \multirow{2}{*}{FLOPs\tnote{1}} & AS-2M & AS-20K & ESC-50 \\
~of SBLU &  & mAP(\%) & mAP(\%) & Acc(\%)\\
\midrule
$K=31$ & 2.41M & 49.8${\scriptstyle \pm 0.0}$ & 41.9${\scriptstyle \pm 0.1}$ & 97.4${\scriptstyle \pm 0.1}$ \\
\rowcolor{gray!15}
$K=63$\tnote{2} & 4.90M & 49.8${\scriptstyle \pm 0.0}$ & 41.9${\scriptstyle \pm 0.1}$ & 97.5${\scriptstyle \pm 0.1}$ \\
$K=127$ & 9.86M & \textbf{49.9}${\scriptstyle \pm 0.1}$ & \textbf{42.0}${\scriptstyle \pm 0.0}$ & \textbf{97.6}${\scriptstyle \pm 0.1}$ \\
\bottomrule
\end{tabular}
\begin{tablenotes}
        \footnotesize
        \item[1] FLOPs are computed only for the adaptive depthwise convolution within SBLU (forward pass only).
        \item[2] Gray-shaded row denotes the default configuration.
\end{tablenotes}
\end{threeparttable}
\label{tab:exp_ablation_kernel_size}
\end{table}
\textbf{Kernel Size of SBLU.}\quad Table~\ref{tab:exp_ablation_kernel_size} examines the effect of kernel size $K$ in Adaptive SBLU, generating a complex sinusoidal kernel modulated by a two-sided exponential window, and reports FLOPs for the adaptive depthwise convolution only.
Relative to the default $K=63$, increasing $K$ to $127$ yields uniform $+0.1$ absolute improvements on AS-2M mAP, AS-20K mAP, and ESC-50 accuracy, while increasing compute from 4.90M to 9.86M FLOPs; decreasing $K$ to $31$ preserves mAP on AS-2M and AS-20K and reduces ESC-50 by $-0.1$, with 2.41M FLOPs.
Overall, we observe a mild positive trend whereby broader receptive fields (larger $K$) tend to yield slightly higher scores across the three benchmarks.

\setlength{\tabcolsep}{3.2pt}
\begin{table}[t]
        \caption{Ablation of Predictor Architecture for Fine-tuning. }
\centering
\begin{threeparttable}
\begin{tabular}{c|ccccc}
\toprule
\multirow{2}{*}{Predictor Arch.} & \multirow{2}{*}{\#Params.\tnote{1}} & Throughput\tnote{2} & AS-2M & AS-20K & ESC-50 \\
& & (batch/sec) & mAP(\%) & mAP(\%) & Acc(\%)\\
\midrule
CNN & ~2.59M & 2.72 & \textbf{49.9}${\scriptstyle \pm 0.1}$ & 41.8${\scriptstyle \pm 0.1}$ & 97.4${\scriptstyle \pm 0.1}$ \\
\rowcolor{gray!15}
Cross-Attention\tnote{3} & 10.63M & \textbf{2.83} & 49.8${\scriptstyle \pm 0.0}$ & \textbf{41.9}${\scriptstyle \pm 0.1}$ & \textbf{97.5}${\scriptstyle \pm 0.1}$ \\
\bottomrule
\end{tabular}
\begin{tablenotes}
        \footnotesize
        \item[1] Parameter counts report predictor parameters only.
        \item[2] Throughput denotes end-to-end training throughput (forward and backward passes).
        \item[3] Gray-shaded row denotes the default configuration.
\end{tablenotes}
\end{threeparttable}
\label{tab:exp_ablation_arch}
\end{table}
\textbf{Predictor Architecture.}\quad In our method, the mask predictor is instantiated as a cross-attention module that uses the mask tokens as queries and visible tokens as keys/values.
Table~\ref{tab:exp_ablation_arch} reports fine-tuning results when replacing this predictor with the CNN used in prior work~\cite{eat-24, sslam-25, asda-25}.
Although the cross-attention predictor has more parameters than the CNN, it delivers higher end-to-end throughput and essentially comparable downstream accuracy with small, consistent gains on two datasets.
Relative to the CNN, throughput increases from 2.72 to 2.83 batches/sec, while the downstream differences are $-0.1$ mAP on AS-2M, $+0.1$ mAP on AS-20K, and $+0.1$ percentage points on ESC-50, respectively.
We hypothesize two contributing factors: following~\cite{crossmae-24}, 25\% of mask tokens are randomly dropped during training, reducing computation, and optimized attention implementations are employed.
These observations suggest that cross-attention offers a favorable accuracy-efficiency trade-off for the mask predictor under our training setup.

\setlength{\tabcolsep}{8.0pt}
\begin{table}[t]
        \caption{Ablation of Clip-level Loss Function for Fine-tuning.}
\centering
\begin{threeparttable}
\begin{tabular}{c|ccc}
\toprule
\multirow{2}{*}{Clip-level Loss} & AS-2M & AS-20K & ESC-50 \\
& mAP(\%) & mAP(\%) & Acc(\%)\\
\midrule
Disabled\tnote{1}~~($\eta_u=0$) & 49.1${\scriptstyle \pm 0.0}$ & 40.2${\scriptstyle \pm 0.1}$ & 97.2${\scriptstyle \pm 0.1}$ \\
\rowcolor{gray!15}
Enabled\tnote{2}~~~($\eta_u=1$) & \textbf{49.8}${\scriptstyle \pm 0.0}$ & \textbf{41.9}${\scriptstyle \pm 0.1}$ & \textbf{97.5}${\scriptstyle \pm 0.1}$ \\
\bottomrule
\end{tabular}
\begin{tablenotes}
        \footnotesize
        \item[1] In the disabled setting, the clip-level loss is removed and the class token is not used during pre-training/fine-tuning.
        \item[2] Gray-shaded row denotes the default configuration.
\end{tablenotes}
\end{threeparttable}
\label{tab:exp_ablation_clip_loss}
\end{table}
\textbf{Clip-level Loss.}\quad The EAT-style teacher-student framework includes a clip-level alignment loss $\mathcal{L}_u$ that matches the student’s class token to a mean-pooled teacher target (Sec.~\ref{sec:loss}).
To examine the contribution of this clip-level objective under our setup, we ablate it by setting $\eta_u=0$ and removing the class token during pre-training/fine-tuning.
As shown in Table~\ref{tab:exp_ablation_clip_loss}, disabling $\mathcal{L}_u$ consistently degrades fine-tuning performance across AS-2M, AS-20K, and ESC-50, indicating that the clip-level objective provides complementary global supervision on top of patch-level masked prediction and improves downstream transfer in our framework.

\setlength{\tabcolsep}{3pt}
\begin{table}[t]
\begin{threeparttable}
\caption{Representation drift under circular temporal shifts on the AS-20K test split (mean $\pm$ std. across test samples).}
\label{tab:exp_shift_consistency}
\centering
\begin{tabular}{cc|cccc|c}
\toprule
\multirow{2}{*}{Framework} & Patch & \multicolumn{5}{c}{Representation drift ($\times 10^{-3}$)} \\
& Embedding & 10 ms & 20 ms & 40 ms & 80 ms & 160 ms\tnote{2} \\
\midrule
EAT  & SPE  & \textbf{0.7}${\scriptstyle \pm 2.7}$ & 1.3${\scriptstyle \pm 4.9}$ & 2.4${\scriptstyle \pm 8.5}$ & 3.6${\scriptstyle \pm 12.6}$ & 1.0${\scriptstyle \pm 2.6}$ \\
EAT  & AaPE & 1.1${\scriptstyle \pm 3.9}$ & 2.1${\scriptstyle \pm 8.1}$ & 3.3${\scriptstyle \pm 10.6}$ & 3.8${\scriptstyle \pm 11.3}$ & 1.7${\scriptstyle \pm 8.4}$ \\
AaSP & SPE  & 1.0${\scriptstyle \pm 2.3}$ & 1.5${\scriptstyle \pm 2.8}$ & 2.0${\scriptstyle \pm 2.7}$ & 2.8${\scriptstyle \pm 3.7}$ & 2.2${\scriptstyle \pm 4.3}$ \\
AaSP & AaPE & \textbf{0.7}${\scriptstyle \pm 1.6}$ & \textbf{1.0}${\scriptstyle \pm 1.8}$ & \textbf{1.5}${\scriptstyle \pm 1.9}$ & \textbf{1.9}${\scriptstyle \pm 1.9}$ & 1.4${\scriptstyle \pm 2.6}$ \\
\bottomrule
\end{tabular}
\begin{tablenotes}
\footnotesize
\item[1] For a fair comparison between SPE and AaPE within each framework, we follow the matched SPE-vs.-AaPE ablation setting and use the best-performing SPE configuration. Accordingly, EAT uses no contrastive regularization, whereas AaSP uses $\eta_c=0.01$.
\item[2] A patch-aligned shift of 160 ms corresponds to a full patch shift under our default patch configuration ($P_\text{time}=16$ with a 10 ms hop size).
\end{tablenotes}
\end{threeparttable}
\end{table}
\subsection{Shift-Consistency Analysis}
To examine whether the proposed framework yields representations that are more stable under small temporal perturbations relevant to aliasing, we evaluate representation drift under circular temporal shifts.
The motivation is that temporal patchification with stride $S_{\text{time}} > 1$ effectively downsamples subband trajectories, and any aliasing introduced in such a representation can manifest as sensitivity to small time shifts or phase offsets.
If the learned representation is more robust to aliasing-sensitive perturbations, the representation of a slightly shifted version of the same audio should remain more consistent.
We use frozen models pre-trained on AS-2M and evaluate on the AS-20K test split.
For each input waveform $x$, we generate circularly shifted versions $x^\delta$ with shift amounts $\delta \in \{10, 20, 40, 80, 160\}\,\mathrm{ms}$.
We then compute the representation drift as $1 - \cos(\mathbf{z}(x), \mathbf{z}(x^\delta)),$ where $\mathbf{z}(\cdot)$ denotes the frozen class token representation.
Smaller values indicate greater consistency under temporal shifts.

Table~\ref{tab:exp_shift_consistency} reports the results.
Within the EAT framework (without contrastive regularization), replacing SPE with AaPE does not improve shift consistency consistently.
In contrast, within the full AaSP framework, AaPE yields lower drift than SPE across all tested shift magnitudes.
Together with the matched SPE-vs-AaPE comparison under the AaSP learning design (Table~\ref{tab:exp_ablation_pe_loss}) and the replacement-only ablation in EAT (Table~\ref{tab:exp_ablation_patch_embedding_replacement}), these results are consistent with our interpretation that AaPE-derived features are not reliably exploited in isolation, but become more useful when coupled with a consistency-promoting learning design.
We stress that this analysis does not demonstrate that aliasing is eliminated or directly quantify aliasing itself; rather, it provides indirect evidence that the full AaSP framework learns representations that are more stable under small temporal perturbations that are relevant to the aliasing issue considered in this work.

\begin{figure}[t]
        \centering
        \includegraphics[width=\linewidth]{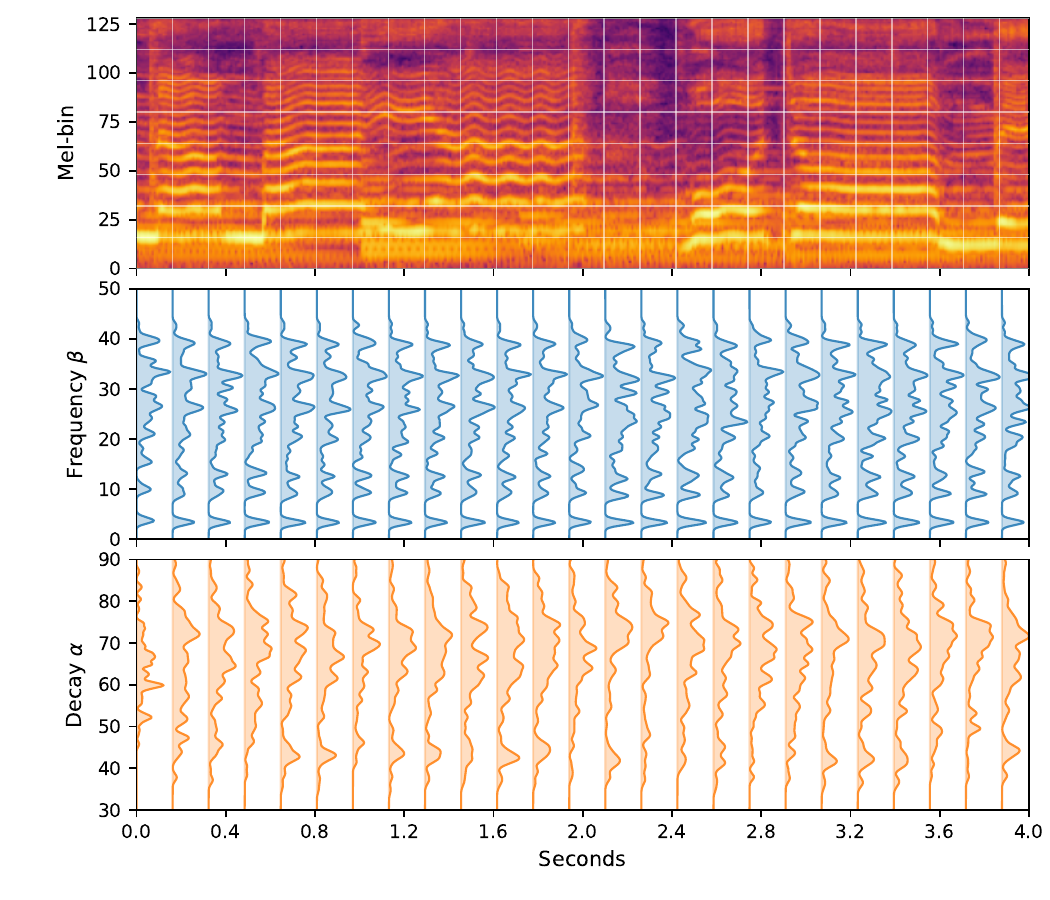}
        \caption{\textbf{Qualitative Example of Patch-wise Adaptive Estimation of SBLU Kernel Parameters.} Top: input spectrogram prior to patchification, with the patch grid overlaid. Middle: estimated decay values for each time patch. Bottom: estimated frequency values for each time patch. The figure shows input-dependent parameter variation for one example.}
        \label{fig:analysis}
\end{figure}
\subsection{Qualitative Interpretability Analysis}
We analyze the behavior of the pre-trained Adaptive SBLU, focusing on its estimations of the frequency and decay.
Figure~\ref{fig:analysis} illustrates the input spectrogram of one sample from AudioSet~\cite{audioset-17}, together with the frequency and decay values estimated by the pre-trained Lambda Encoder for that input.
In our experimental setup, Adaptive SBLU estimates 16 pairs of frequency and decay for each patch.
This figure is intended as a qualitative case study of input-dependent parameter variation, rather than as a quantitative analysis of the dataset-level distribution of the estimated parameters.
For this example, the estimated frequency values are spread across the target range, which is consistent with the intended role of Adaptive SBLU in analyzing cues from alias-prone modulation bands within $[\beta_{\min}, \beta_{\max}]$.
The decay values are often skewed toward larger values, while lower values also appear at some time points, illustrating that the estimated time-frequency trade-off can vary with the input.
Overall, the figure qualitatively suggests that Adaptive SBLU can vary its analysis parameters across time patches in an input-dependent manner.

\section{Conclusion}
This work investigates aliasing-sensitive representation instability that can arise from spectrogram patchification in transformer-based audio self-supervised learning and proposes a framework designed to learn more stable representations under such perturbations.
We propose AaSP, an aliasing-aware self-supervised pre-training framework whose key component is Aliasing-aware Patch Embedding (AaPE), implemented using the Structured Bilateral Laplace Unit (SBLU).
Within this framework, SBLU employs a two-sided exponential window for dynamic subband frequency analysis, and the resulting features are fused with standard patch tokens through AaPE.
This approach aims to incorporate information from alias-prone modulation bands without resorting to blanket low-pass filtering that may discard task-relevant high-frequency content.
Within AaSP, AaPE is combined with teacher-student masked audio modeling, a multi-mask training scheme, and contrastive regularization.
Our ablations show that AaPE is not a universally effective standalone replacement; rather, its benefit emerges when it is paired with a learning design that encourages consistency across masked views.

AaSP achieves strong fine-tuning performance across diverse audio benchmarks, with particularly clear gains on several general acoustic and music-related tasks, while remaining competitive under linear evaluation.
These trends suggest that frequency-sensitive feature extraction can be effective for periodic, quasi-stationary audio.
However, tasks dominated by fine-grained pitch dynamics or significantly short contexts may require complementary mechanisms.
Because AaPE targets the input representation stage, it may be complementary to other SSL improvements, although validating such combinations remains future work.

Our current interpretation is that AaPE exposes features from modulation bands that are vulnerable under strided patchification.
However, these features are not reliably exploited under masked modeling when AaPE is used in isolation.
In the full AaSP framework, multi-mask training and contrastive regularization encourage different masked views of the same audio to map to consistent representations, which appears to help the model use AaPE-derived features in a more stable way.
This interpretation is supported by the replacement-only ablation and by the shift-consistency analysis, although the exact causal contribution of each component remains an important topic for future work.

The limitations of this approach include the reliance on log-mel inputs, emphasis on amplitude-modulation cues, and reduced gains on extremely short utterances or pitch-dominated tasks.
Future work will combine AaSP with stronger pre-training strategies, evaluate AaSP on automatic speech recognition (ASR) benchmarks using standard ASR fine-tuning and decoding pipelines, and extend SBLU to better capture pitch-varying/chirp patterns and very short contexts.

\bibliographystyle{IEEEtran}
\bibliography{references}

\appendices

\section{} \label{sec:appendix_a}
\renewcommand{\theequation}{A.\arabic{equation}}
\setcounter{equation}{0}
We derive Eq. (\ref{eq:sblu_complex}) under the notation and conventions specified in the main text (see Sec.\ref{sec:sblu}).
Applying the integrating factor $e^{-\mathbf{A}t}$ to the continuous-time linear state equation $\frac{d}{dt}\mathbf{h}(t)=\mathbf{A}\mathbf{h}(t)+\mathbf{B}\mathbf{u}(t)$ and integrating from $0$ to $s$ yields
\begin{equation}
        \mathbf{h}(s) = e^{\mathbf{A}s}\mathbf{h}(0)+ \int_0^s e^{\mathbf{A}(s-t)}\mathbf{B}\mathbf{u}(t)\,dt,
\end{equation}
with the initial condition $\mathbf{h}(0)=0$.
As a modeling choice to obtain a symmetric representation (two-sided exponential window) centered at $s/2$, we write
\begin{equation}
        \mathbf{h}(s) = \int_0^s e^{-\mathbf{A}\left|t-\frac{s}{2}\right|}\mathbf{B}\mathbf{u}(t)\,dt,
\end{equation}
which, upon splitting at $s/2$, becomes
\begin{equation}
        \mathbf{h}(s) = e^{-\frac{\mathbf{A}s}{2}}\int_0^\frac{s}{2} e^{\mathbf{A}t}\mathbf{B}\mathbf{u}(t)\,dt + e^{\frac{\mathbf{A}s}{2}}\int_\frac{s}{2}^s e^{-\mathbf{A}t}\mathbf{B}\mathbf{u}(t)\,dt.
\end{equation}
Let $s = (n+1)\Delta$ with $n=K-1$, and apply zero-order-hold (ZOH) discretization.
We obtain
\begin{align}
        \mathbf{h}[K]
&= e^{-\frac{1}{2}\Delta\mathbf{A}(n+1)} \sum_{k=0}^{c}  \left( \int_{k\Delta}^{(k+1)\Delta}e^{\mathbf{A}t}\,dt \right)\mathbf{B}\mathbf{u}[k] \notag\\
&\quad+ e^{\frac{1}{2}\Delta\mathbf{A}(n+1)} \sum_{k=c+1}^{K-1} \left( \int_{k\Delta}^{(k+1)\Delta}e^{-\mathbf{A}t}\,dt  \right)\mathbf{B}\mathbf{u}[k] \notag\\
&= e^{-\frac{1}{2}\Delta\mathbf{A}}\mathbf{A}^{-1}(e^{\Delta\mathbf{A}} - \mathbf{I})\sum_{k=0}^{c} e^{\Delta\mathbf{A}(k-c)}\mathbf{B}\mathbf{u}[k] \notag\\
&\quad+ e^{\frac{1}{2}\Delta\mathbf{A}}\mathbf{A}^{-1}(\mathbf{I} - e^{-\Delta\mathbf{A}})\sum_{k=c+1}^{K-1} e^{\Delta\mathbf{A}(c-k)}\mathbf{B}\mathbf{u}[k],
\end{align}
using $\int_{k\Delta}^{(k+1)\Delta} e^{\pm \mathbf{A}t}\,dt = \pm\mathbf{A}^{-1}e^{\pm \Delta\mathbf{A}k}(e^{\pm \Delta\mathbf{A}}-\mathbf{I})$.
With the eigen-decomposition and identities $\mathbf{A} = \mathbf{P}\mathbf{\Lambda} \mathbf{P}^{-1}, e^\mathbf{A} = \mathbf{P}e^{\mathbf{\Lambda}}\mathbf{P}^{-1}, \sinh\left(\tfrac{1}{2}\Delta\mathbf{\Lambda}\right)=\tfrac{1}{2}\left(e^{\frac{1}{2}\Delta\mathbf{\Lambda}}-e^{-\frac{1}{2}\Delta\mathbf{\Lambda}}\right),$ we obtain
\begin{align}
        \mathbf{h}[K]
&= \mathbf{P}e^{-\frac{1}{2}\Delta\mathbf{\Lambda}}\mathbf{\Lambda}^{-1}(e^{\Delta\mathbf{\Lambda}} - \mathbf{I})\sum_{k=0}^{c} e^{\Delta\mathbf{\Lambda}(k-c)}\mathbf{P}^{-1}\mathbf{B}\mathbf{u}[k] \notag\\
&+ \mathbf{P}e^{\frac{1}{2}\Delta\mathbf{\Lambda}}\mathbf{\Lambda}^{-1}(\mathbf{I} - e^{-\Delta\mathbf{\Lambda}})\sum_{k=c+1}^{K-1} e^{\Delta\mathbf{\Lambda}(c-k)}\mathbf{P}^{-1}\mathbf{B}\mathbf{u}[k] \notag\\
&= 2\mathbf{P}\mathbf{\Lambda}^{-1}\sinh\left(\frac{\Delta\mathbf{\Lambda} }{2}\right) \sum_{k=0}^{K-1} e^{-\Delta\mathbf{\Lambda} |k - c|} \overline{\mathbf{B}}\mathbf{u}[k],
\end{align}
where $\overline{\mathbf{B}}=\mathbf{P}^{-1}\mathbf{B}$.
Finally, regarding $\mathbf{h}[K]$ as the representative value at the center index of the $n$-th sliding window, and applying the observation equation $\mathbf{y}[n] = \mathbf{C}\mathbf{h}[n]$, we obtain
\begin{equation}
        \mathbf{y}[n] = 2\overline{\mathbf{C}} \mathbf{\Lambda}^{-1} \sinh \left( \frac{\Delta\mathbf{\Lambda}}{2}\right)\sum_{k=0}^{K-1} e^{-\Delta\mathbf{\Lambda} \left| k - c \right|} \overline{\mathbf{B}} \mathbf{u}[n+k-c],
\end{equation}
with $\overline{\mathbf{C}}=\mathbf{C}\mathbf{P}$.

\section{} \label{sec:appendix_b}
\renewcommand{\theequation}{B.\arabic{equation}}
\setcounter{equation}{0}
We derive the normalization coefficient $\gamma_i \geq 0$ in Eq. (\ref{eq:sblu_gamma}).
Let $\lambda_i=\alpha_i + j\beta_i$ with $\alpha_i,\beta_i\in\mathbb{R}$ and $j^2=-1$.
For the diagonal elements of the complex diagonal matrix $\mathbf{\Lambda}^{-1} \sinh \left( \frac{\Delta\mathbf{\Lambda}}{2}\right)$, namely $\lambda_i^{-1}\sinh\left(\frac{\Delta\lambda_i}{2}\right)$, employing the identity for a complex argument of $\sinh(\cdot)$ (and Euler’s formula) yields
\begin{align}
        \sinh\left(\frac{\Delta\lambda_i}{2}\right)
&= \sinh\left(\frac{\Delta\alpha_i}{2}\right)\cos\left(\frac{\Delta\beta_i}{2}\right) \notag \\
&\quad+ j\cosh\left(\frac{\Delta\alpha_i}{2}\right)\sin\left(\frac{\Delta\beta_i}{2}\right).
\end{align}
Using $\cosh^2(\cdot)-\sinh^2(\cdot)=1$, we obtain
\begin{equation}
        \gamma_i = \left|\lambda_i^{-1}\sinh\left(\frac{\Delta\lambda_i}{2}\right)\right|
= \sqrt{\frac{\cosh^2\!\left(\frac{\Delta\alpha_i}{2}\right) - \cos^2\!\left(\frac{\Delta\beta_i} {2}\right)}{\alpha_i^2 + \beta_i^2}}.
\end{equation}
Equivalently, noting that for $a,b\in\mathbb{R}$, $\left|\sinh(a+jb)\right|^2 = \sinh^2 a \cos^2 b + \cosh^2 a \sin^2 b = \cosh^2 a - \cos^2 b$, the non-negativity $\gamma_i\ge 0$ follows immediately, and the denominator is positive since $|\lambda_i| > 0$ by the imposed bounding constraints.

\section{} \label{sec:appendix_c}
\renewcommand{\theequation}{C.\arabic{equation}}
\setcounter{equation}{0}
We illustrate, in a scalar example, the coupling that enables convolution reuse in Adaptive SBLU (see Sec. \ref{sec:aape}): the real-part gradient with respect to the decay $\alpha$ and the imaginary-part gradient with respect to the frequency $\beta$ share an identical convolution; likewise, the real-part gradient with respect to $\beta$ and the imaginary-part gradient with respect to $\alpha$ share an identical convolution.
For an input $x_n \in \mathbb{R}$ at time index $n$ with parameters $\alpha_n$ (decay) and $\beta_n$ (frequency), define the magnitude output $|y_n|=\sqrt{\text{Re}(y_n)^2 + \text{Im}(y_n)^2}$.
The real and imaginary parts are
\begin{align}
\text{Re}(y_n) &= \sum_{k=0}^{K-1} x_{n+k-c} \cdot e^{-\Delta_k\alpha_n} \cos(\Delta_k\beta_n) \\
\text{Im}(y_n) &= \sum_{k=0}^{K-1} x_{n+k-c} \cdot e^{-\Delta_k\alpha_n} \sin(\Delta_k\beta_n)
\end{align}
with $\Delta_k = \Delta |k - c|$.
Setting $R_n = \frac{\text{Re}(y_n)}{|y_n|}$ and $I_n = \frac{\text{Im}(y_n)}{|y_n|}$, the real-part gradients of $|y_n|$ are
\begin{align}
\text{Re}\left( \frac{\partial |y_n|}{\partial \alpha_n} \right)
&= - R_n \sum_{k=0}^{K-1} \Delta_k\, x_{n+k-c} e^{-\Delta_k\alpha_n} \cos(\Delta_k\beta_n) \\
\text{Re}\left(\frac{\partial |y_n|}{\partial \beta_n} \right)
&= - R_n \sum_{k=0}^{K-1} \Delta_k\, x_{n+k-c} e^{-\Delta_k\alpha_n} \sin(\Delta_k\beta_n)
\end{align}
and the imaginary-part gradients are
\begin{align}
\text{Im}\left( \frac{\partial |y_n|}{\partial \alpha_n} \right)
&= -I_n \sum_{k=0}^{K-1} \Delta_k\, x_{n+k-c} e^{-\Delta_k\alpha_n} \sin(\Delta_k\beta_n) \\
\text{Im}\left( \frac{\partial |y_n|}{\partial \beta_n} \right)
&= I_n \sum_{k=0}^{K-1} \Delta_k\, x_{n+k-c} e^{-\Delta_k\alpha_n} \cos(\Delta_k\beta_n)
\end{align}
from which the following coupling relations follow, halving the convolutional workload:
\begin{align}
\text{Re}\left( \frac{\partial |y_n|}{\partial \alpha_n} \right) &= -\frac{R_n}{I_n} \text{Im}\left(\frac{\partial |y_n|}{\partial \beta_n} \right), \\
\text{Re}\left(\frac{\partial |y_n|}{\partial \beta_n} \right) &= \frac{R_n}{I_n} \text{Im}\left(\frac{\partial |y_n|}{\partial \alpha_n} \right).
\end{align}
Finally, the gradients are
\begin{align}
\frac{\partial |y_n|}{\partial \alpha_n}
&= \text{Re}\left(\frac{\partial |y_n|}{\partial \alpha_n} \right) +  \text{Im}\left(\frac{\partial |y_n|}{\partial \alpha_n} \right) \notag\\
&= \text{Re}\left(\frac{\partial |y_n|}{\partial \alpha_n} \right) + \frac{I_n}{R_n} \text{Re}\left(\frac{\partial |y_n|}{\partial \beta_n} \right), \\
\frac{\partial |y_n|}{\partial \beta_n}
&= \text{Re}\left(\frac{\partial |y_n|}{\partial \beta_n} \right) +  \text{Im}\left(\frac{\partial |y_n|}{\partial \beta_n} \right) \notag\\
&= \text{Re}\left(\frac{\partial |y_n|}{\partial \beta_n} \right) - \frac{I_n}{R_n}\text{Re}\left( \frac{\partial |y_n|}{\partial \alpha_n} \right).
\end{align}
In implementation, it suffices to compute once per $n$ the two base convolutions
\begin{align}
S_c(n)&=\sum_{k=0}^{K-1}\Delta_k\, x_{n+k-c}\, e^{-\Delta_k\alpha_n} \cos(\Delta_k\beta_n),\\
S_s(n)&=\sum_{k=0}^{K-1}\Delta_k\, x_{n+k-c}\, e^{-\Delta_k\alpha_n} \sin(\Delta_k\beta_n),
\end{align}
and reuse them across the four partial gradients via the coupling relations above, reducing compute and memory traffic.

\section{} \label{sec:appendix_flops}
\renewcommand{\theequation}{D.\arabic{equation}}
\setcounter{equation}{0}
We provide an approximate FLOPs estimate for the patch-embedding stage under the default setting ($F{=}128$, $T{=}608$, $P_\text{freq}{=}P_\text{time}{=}16$, $D{=}768$, $K{=}63$, $H{=}128$, $C{=}1024$).
We count one MAC as 2 FLOPs and ignore minor operations.
Let $\widetilde{F}=F/P_\text{freq}=8$, $\widetilde{T}=T/P_\text{time}=38$, and $N=\widetilde{F}\widetilde{T}=304$.

\textbf{SPE.} A single strided Conv2D with kernel $(16{\times}16)$ and stride $(16,16)$ yields
\begin{equation}
{(\mathrm{FLOPs})}_\mathrm{SPE} \approx 2 \cdot N \cdot D \cdot 16 \cdot 16 \approx 0.12~\mathrm{GFLOPs}.
\end{equation}

\textbf{AaPE.} We report the AaPE estimate for the masked-input pre-training regime (80\% mask ratio), where the Lambda Encoder processes $N_\mathrm{vis}\approx 0.2N$ visible tokens.
We approximate the total cost as the sum of the dominant terms:
(i) Patch Fusion, a token-wise linear map $(D{+}\widetilde{C})\!\to\!D$ with $\widetilde{C}=C/\widetilde{F}=128$,
\begin{equation}
{(\mathrm{FLOPs})}_\mathrm{fuse} \approx 2 \cdot N \cdot D \cdot (D+\widetilde{C});
\end{equation}
(ii) Adaptive SBLU, which consists of the zero-phase high-pass filtering step, a length-$K$ depthwise 1D convolution over $H$ channels, and grouped $1{\times}1$ projections.
Using a 63-tap FIR implemented by forward--backward filtering (two passes), we approximate
\begin{equation}
        {(\mathrm{FLOPs})}_\mathrm{HPF} \approx 4 \cdot F \cdot T \cdot 63,
\end{equation}
and write the dominant Adaptive SBLU cost as
\begin{align}
{(\mathrm{FLOPs})}_\mathrm{SBLU} &\approx {(\mathrm{FLOPs})}_\mathrm{HPF} \nonumber\\
&+\; 2 \cdot H \cdot T \cdot K \;+\; 2 \cdot T \left(HF + CH\right);
\end{align}
and (iii) the Lambda Encoder applied to $N_\mathrm{vis}$ tokens (we approximate it using standard Transformer-block accounting at width $D/r$ with $r{=}6$ and $L{=}3$ blocks).
Under these assumptions, the AaPE patch-embedding cost is approximately 0.55 GFLOPs per clip for a forward pass.

\end{document}